\documentclass[11pt]{academia}  
\pdfoutput=1
\usepackage{graphicx}
\usepackage{color}
\usepackage{amssymb}
\usepackage{amsmath}
\usepackage{relsize}
\usepackage{mathabx}
\usepackage{upgreek}
\usepackage{bbm}
\usepackage{multirow}
\usepackage{multicol}
\usepackage{amscd}
\usepackage{ipa}
\usepackage{mfpic}
\usepackage{verbatim}
\usepackage{cancel}
\usepackage{chemarrow}
\usepackage{tikz-cd}
\usepackage{tocloft}

\usepackage[linktocpage=false,colorlinks = true,
            linkcolor = darkviolet,
            urlcolor  = blue,
            citecolor = blue,
            anchorcolor = magenta]{hyperref} 

\definecolor{cream}{rgb}{.97, .95, .88}
\definecolor{darkcream}{rgb}{1., .88, .5}
\definecolor{lightpink}{rgb}{0.98, 0.88, 0.87}
\definecolor{lightwhite}{rgb}{1., 0.98, 0.95}
\definecolor{lightsalmon}{rgb}{1., 0.95, 0.90}
\definecolor{lightviolet}{rgb}{0.9, 0.8, 0.9}
\definecolor{lightgray}{rgb}{.96, .96, .96}  
\definecolor{lgray}{rgb}{.75, .75, .75}
\definecolor{LemonChiffon}{rgb}{0.95, 1., 0.7}
\definecolor{lightolivegreen}{rgb}{0.84, 0.89, 0.25}
\definecolor{lightgreen}{rgb}{.664, 1., .52}
\definecolor{llgreen}{rgb}{.900, .983, .960}
\definecolor{tristle}{rgb}{0.87, 0.67, 0.77} 
\definecolor{pink}{rgb}{0.95, 0.45, 0.75}
\definecolor{magenta}{rgb}{1., 0, 1.}
\definecolor{violet}{rgb}{0.9, 0.20, 0.85}
\definecolor{darkolivegreen}{rgb}{0.55, 0.65, 0.35}
\definecolor{maroon}{rgb}{0.7, 0.26, 0.56}
\definecolor{lightmaroon}{rgb}{0.85, 0.38, 0.58}
\definecolor{darkmaroon}{rgb}{0.604, 0.169, 0.451}
\definecolor{ddarkmaroon}{rgb}{0.2, 0.03125, 0.150}
\definecolor{mediumorchid}{rgb}{0.8, 0.33, 0.83}
\definecolor{mediumorchidd}{rgb}{1., 0.33, 0.63}
\definecolor{darkgreen}{rgb}{0.1, 0.6, 0.13}
\definecolor{lightyellow}{rgb}{1., 1., 0.82}
\definecolor{turquoise}{rgb}{0.042, 0.586, 0.512}
\definecolor{turquoisel}{rgb}{0.66, 0.94, 0.83}
\definecolor{darkturquoise}{rgb}{0.21, 0.55, 0.50}
\definecolor{coral}{rgb}{1., 0.6, 0.21}
\definecolor{lightorange}{rgb}{1., 0.88, 0.75}
\definecolor{orangered}{rgb}{1., 0.5, 0.}
\definecolor{orange}{rgb}{1., 0.65, 0.1}
\definecolor{orangel}{rgb}{1., .85, .3}
\definecolor{darkorange}{rgb}{0.875, 0.4, 0.204}
\definecolor{ddarkorange}{rgb}{.675, .218, .05}
\definecolor{bluesky}{rgb}{0.48, 0.53, 1.}
\definecolor{gold}{rgb}{1., 0.85, 0.25}
\definecolor{goldd}{rgb}{0.95, 0.75, 0.05}
\definecolor{darkviolet}{rgb}{0.54, 0.04, 0.84}
\definecolor{ddarkviolet}{rgb}{.382, .063, .657}
\definecolor{lightblue}{rgb}{0.30, 0.86, 0.89}
\definecolor{LightBlue}{rgb}{0.68, 0.85, 0.9}
\definecolor{lblue}{rgb}{0.78, 0.90, 0.95}
\definecolor{darkblue}{rgb}{.105, .308, .707}
\definecolor{lightmaroon}{rgb}{0.85, 0.38, 0.58}
\definecolor{darkmaroon}{rgb}{0.604, 0.169, 0.451}
\definecolor{darkpink}{rgb}{0.879, 0.020, 0.766}
\definecolor{ddarkpink}{rgb}{0.738, 0.195, 0.406}
\definecolor{grey}{rgb}{0.717, 0.717, 0.717}
\definecolor{lightgrey}{rgb}{0.800, 0.800, 0.800}
\definecolor{brown}{rgb}{0.740, 0.323, 0.182}
\definecolor{redbrown}{rgb}{.575, .158, .05}
\definecolor{darkbrown}{rgb}{0.34, 0.25, 0.05}
\definecolor{orangebrown}{rgb}{0.433, 0.262, 0.06}
\definecolor{pinkl}{rgb}{1., 0.788, 0.918}
\definecolor{salmon}{rgb}{1., 0.66, 0.5}
\definecolor{lightbrown}{rgb}{0.703, 0.508, 0.121}

\def\Journal#1#2#3#4{{#1} {\bf #2} (#3) #4}

\def\etal{{\it et al.}}
\def\Name#1#2 {{ #1 }{#2}}

\def\APH{\em Annals Phys.}

\def\ATM{\em Adv. Theor. Math. Phys}

\def\CMP{\em Commun. Math. Phys.}

\def\CQG{\em Class. Quant.~Grav.}

\def\FOP{\em Found. Phys.}

\def\GRG{\em Gen. Rel. Grav}

\def\IMA{{\em Int. J. Mod. Phys.} \emph{A}}
\def\IMD{{\em Int. J. Mod. Phys.} \emph{D}}

\def\JHE{\em J. High Energy Phys.}

\def\JMP{\em J. Math. Phys.}

\def\JPU{\em J. Phys. (USSR)}

\def\MDU{\em MDPI Universe J.}

\def\MPL{{\em Mod. Phys. Lett.} \emph{A}}

\def\NAC{\em Nature Commun.}

\def\NAT{\em Nature}

\def\NPB{{\em Nucl.~Phys.}~\emph{B}}

\def\PAS{\em Publ. Astro. Soc. Japan}

\def\PLA{{\em Phys. Lett.}~\emph{A}}
\def\PLB{{\em Phys. Lett.}~\emph{B}}

\def\PRA{{\em Phys.~Rev.}~\emph{A}}
\def\PRB{{\em Phys.~Rev.}~\emph{B}}

\def\PRD{{\em Phys.~Rev.}~\emph{D}}

\def\PRL{\em Phys. Rev. Lett.}

\def\PRX{{\em Phys.~Rev.}~\emph{X}}

\def\RMP{\em Rep. Mod. Phys.}

\def\RVM{\em Rev. Math. Phys.}

\def\SYM{\em Symmetry}

\def\be{\begin{equation}}
\def\ee{\end{equation}}
\def\bea{\begin{eqnarray}}
\def\eea{\end{eqnarray}}
\def\bes{\begin{equation*}}
\def\ees{\end{equation*}}
\def\beas{\begin{eqnarray*}}
\def\eeas{\end{eqnarray*}}

\def\tr{\text{tr}}
\def\diag{\text{diag}}

\def\mg{\mathsf g}

\def\cg{\mathtt g}
\def\ch{\mathtt h}
\def\am{\mathcal A}
\def\bm{\mathcal B}

\def\hm{\mathcal H}
\def\lm{\mathcal L}
\def\mm{\mathcal M}
\def\nm{\mathcal N}

\def\um{\mathcal U}

\def\ha{\hat{a}}
\def\hA{\hat{A}}
\def\hB{\hat{B}}

\def\hG{\hat{G}}

\def\hK{\hat{K}}
\def\hL{\hat{L}}
\def\hO{\hat{O}}

\def\hT{\hat{T}}

\def\hrho{\hat{\rho}}
\def\hvarrho{\hat{\varrho}}
\def\hphi{\hat\phi}
\def\sD{\cancel{D}}

\def\sqgr{$SU(\infty)$-QGR}
\def\suinf{{SU(\infty)}}
\def\suinfa{{\mathcal{SU}(\infty)}}

\begin{document}
\articletype{Research}
\proofstage{Preprint}
\jName{Academia Quantum}

\title{Quantum state of fields in $\boldsymbol{\suinf}$ Quantum Gravity}

\author{$^{1*,2*}$Houri Ziaeepour}
\maketitle

\affiliation{$^{1}$Institut UTINAM, CNRS UMR 6213, Observatoire de Besan\c{c}on, Universit\'e de Franche Compt\'e, 41 bis ave. de l'Observatoire, BP 1615, 25010 Besan\c{c}on, France} \\
\affiliation{$^{2}$Mullard Space Science Laboratory, University College London, Holmbury St. Mary, GU5 6NT, Dorking, Surrey, UK}
\email{houriziaeepour@gmail.com}



\newcounter{propos}

\section*{abstract}
Our Universe is ruled by quantum mechanics and should be treated as a quantum system. \sqgr~is 
a recently proposed quantum model for the Universe, in which gravity is associated to $\suinf$ 
symmetry of its Hilbert space. Fragmentation of its infinite dimensional state due to random 
quantum fluctuations divides the Universe to approximately isolated subsystems. In addition to 
parameters of their {\it internal} finite rank symmetries, states and dynamics of subsystems are 
characterized by 4 continuous parameters and the perceived classical spacetime is their effective 
representation, reflecting quantum states of subsystems and their relative evolution. At lowest 
order the effective Lagrangian of \sqgr~has the form of Yang-Mills gauge theories for both 
$\suinf$ - gravity - and internal symmetries defined on the aforementioned 4D parameter space. 
In the present work we study more thoroughly some of the fundamental aspects of \sqgr. 
Specifically, we clarify the impact of the degeneracies of $\suinfa$ algebra on the construction 
of the model; describe mixed states of subsystems and their purification; calculate measures of 
their entanglement to the rest of the Universe; and discuss their role in the emergence of local 
gauge symmetries. We also describe the relationship between what is called {\it internal space} 
of $\suinf$ Yang-Mills with the 4D parameter space, and analytically demonstrate irrelevance of 
the geometry of parameter space for physical observables. Along with these topics, we demonstrate 
the equivalence of two sets of criteria for compositeness of a quantum system, and show 
uniqueness of the limit of various algebras leading to $\suinfa$.

\tableofcontents

\section{Introduction to \sqgr} \label{sec:intro}
\sqgr is a fundamentally quantum approach to cosmology and gravity. Its rationale, foundation 
and essential structure and properties are investigated in~\cite{houriqmsymmgr,hourisqgr}. 
It is reviewed in~\cite{hourisqgrhighlight,hourisqgrshort} and compared with several popular 
Quantum Gravity (QGR) models in~\cite{hourisqgrcomp}. 

We begin by a nutshell description of \sqgr. The essential axiom of this model is the existence 
of infinite number of mutually commuting observables in a quantum Universe. Consequently, the 
Hilbert space $\hm_U$ of the Universe represents $\suinf$ symmetry group~\cite{suninfhoppthesis,suninfvirasoro,suninfvirasoro0,suninfsurfaceanomal,suinftriang,suninftorus,suinftriang0,adifftorussuinf,suninfrep,suninfsimplect,suninfrep0}. There is no background spacetime in this model and the 
usual quantization procedure is replaced by non-commutative $\suinfa$ algebra of linear operators 
$\hO \in \bm[\hm_U]$ acting on $\hm_U$~\cite{qgrnoncommut,qmmathbook}.

It is demonstrated~\cite{houriqmsymmgr} that as a whole this Universe is static and trivial, in 
the sense that any state can be transformed to another one by a $\suinf$ transformation under 
which observables are conserved, Nonetheless, quantum fluctuations, that is random application of 
$\hO \in \bm[\hm_U]$, lead to approximate fragmentation of its state, and thereby the Hilbert 
space $\hm_U$, and an approximate division to subsystems according to the criteria defined 
in~\cite{sysdiv}. 

We remind that the process of Hilbert space fragmentation of closed 
quantum systems to subsystems with their own symmetries and behaviour is observed in many 
condensed matter contexts, see e.g.~\cite{qmfragmentobs} and references therein. Indeed, this 
topic is currently under intensive theoretical~\cite{qmfragalgebra,qmfragsimul} and 
experimental~\cite{qmfragmentobs} investigation, and is reviewed in~\cite{qmfragrev,qmfragrev0}. 
Due to many-body nature of these systems and existence of a large number of degrees of freedom, 
dominance of long range 
electromagnetic interactions, and isolation of the whole system they can be considered as 
{\it mini-universes} which their Hilbert spaces represent $SU(N \rightarrow \infty)$ symmetry. 
Therefore, it is not unreasonable to assume that the Hilbert space $\hm_U$ of the whole 
Universe - the only truly isolated many-body quantum system - represents $\suinf$ symmetry and 
becomes fragmented to subsystems by random quantum fluctuations.

The clustered and fragmented blocks n $\bm[\hm_U]$ represent finite rank {\it internal} 
symmetries, which we collectively call $G$. Moreover, the global $\suinf$ becomes the symmetry 
of their infinite dimensional environment consisting of other subsystems entangled to them. 
Hence, the Hilbert space of a subsystem represents $G \times \suinf$ symmetry. As a consequence, 
representations of $\suinf$ by subsystems become comparable, and a dimensionful area or length 
parameter is added to the list of parameters characterizing quantum state of subsystems. 
In addition, a subsystem can play the role of a quantum clock and define a relative time and 
dynamics \`a la Page and Wootters~\cite{qmtimepage,qmtimepage0}. Mutual entanglement of all 
subsystems, including the clock, with their environment ensures a synchronous relative variation 
and an arrow of time.

Using Mandelstam-Tamm inequality~\cite{qmspeed} - the quantum speed limit - it is shown that a 
classical spacetime emerges as an effective manifestation of the (1+3)-dimensional parameter 
space of the Hilbert space of $\suinf$ symmetry of the subsystems and their relative dynamics. 
Moreover, it is proved that signature of the metric of this emergent spacetime is negative. 
Thus, \sqgr~relates the Lorentzian geometry of classical spacetime to quantum states of the 
content of the Universe and their dynamics. However, \sqgr~does not explore the geometry of the 
Hilbert space\footnote{Geometry has been first introduced in the Hilbert space of many-body 
quantum systems in order to study collective properties, such as deformation of nuclides, 
see~\cite{qmhilbertgeom} and references there in. Considering an Euclidean $\mathbb{R}^n$ 
parameter space for the quantum states, they used variation of state and its path in the 
Hilbert space to define a symmetric 2-tensor as metric and a 2-form as symplectic structure 
interpreted as {\it induced geometry} of parameter space. This geometric approach helps to 
study complicated variations of the quantum state of many-body systems, see 
e.g.~\cite{qmgeomantiferro} for an application example of this method.} per se. Rather, it 
relates the perceived spacetime and its geometry to parameter space of quantum states. It is 
also shown that the geometry of the parameter space is irrelevant for physical observables and 
can be gauged out by a $\suinf$ transformation. 

A functional constructed from invariants of $\suinf$ and $G$ plays the role of action analogous 
to those of Quantum Field Theories (QFT's). At lowest order of symmetry invariant traces 
this effective action is similar to non-Abelian Yang-Mills models for both $\suinf$ and $G$ 
symmetries. There is however an essential difference: the action is defined on the (3+1)D space 
of continuous parameters of the model, rather than a classical spacetime. Due to the Yang-Mills 
nature of interactions, the mediator boson of quantum gravity in \sqgr~is a spin-1 field. 
Nonetheless, it is shown that at classical limit - that is when quantum effects of gravity are 
not detectable - the gravity sector dynamics is ruled by the Einstein-Hilbert action\footnote{The 
non-commutative algebra of $\suinf$ can be normalized such that it depends on $\hbar/M_P$, 
where $M_P$ and $\hbar$ are Planck mass and constant, respectively. This normalization is 
necessary for making pure gravity - $\suinf$ gauge - term of the action to have a dimensionality 
similar to the scalar curvature in the Einstein-Hilbert action. In the standard classical limit 
i.e. $\hbar \rightarrow 0$ or in absence of gravity, i.e. $M_P \rightarrow \infty$ the algebra 
becomes Abelian and the model trivial. \label{foot:classiclim}}.

The aim of present work is to discuss in more details and clarify some of the technical issues 
that were not addressed in the previous works, in particular in~\cite{hourisqgr} where properties 
of \sqgr~summarized in the above paragraphs are studied. Here we only 
remind those details which are necessary for understanding of the content of this work. 
For complementary explanations and demonstrations the reader should refer 
to~\cite{houriqmsymmgr,hourisqgr}. The topics discussed here are necessary both for reinforcing 
the construction and consistency of the model, and for preparing the ground for its applications 
to topics such as black hole information puzzle and prediction of outcomes of future laboratory 
tests of quantum gravity. The main results reported here are summarized in 
Sec. \ref{sec:summary}. 

In Sec. \ref{sec:suinfsqgr}~ we clarify the issue of the degeneracy of 
$\mathcal{SU}(N \rightarrow \infty)$ algebra and its impact on \sqgr's parameter space. 
Emergence of subsystems, their Hilbert spaces, and entanglement of their states with the rest 
of the Universe is investigated in Sec. \ref{sec:subsysstate}. In order to provide explicit 
and quantitative proof of the Proposition 1 in~\cite{hourisqgr}, we quantify the entanglement 
by calculating several of its measures in Appendix \ref {app:entangle}, and compare the results 
with QGR models that try to explain gravity as a consequence of 
entanglement~\cite{qgrentangle,qgrentangle0,qgrentangle1}. In Sec. \ref{sec:dyn}~ we review 
dynamics of the Universe and its subsystems according to \sqgr~and provide an analytical 
description of the relationship between $\suinf$ symmetry of the Hilbert space and 
diffeomorphism of its parameter space. Irrelevance of the geometry of parameter space is the 
subject of Proposition 2 in~\cite{hourisqgr}. Here we investigate and demonstrate this 
crucial property in more details. Finally, in Sec. \ref{sec:outline}~ we 
briefly outline an experiment/observation which may be able to detect signature of \sqgr with 
present facilities. We also discuss topics left for future investigations. Appendices include 
details of arguments, calculations, and reviews of subjects necessary for this work.

\subsection{Summary of main results}  \label{sec:summary}
Regarding \sqgr~model we reinforce demonstration of Propositions 1 and 2 of~\cite{hourisqgr}, 
calculate quantum states of subsystems of the Universe, and investigate their properties:
\begin{description}
\item{\bf Emergence and properties of structures in quantum state of the Universe: } 
In~\cite{hourisqgr} we employed an operational approach to show how structures may emerge in 
the quantum state of the Universe. We demonstrated that they can be approximately treated as 
quantum subsystems. Using those results, here we construct their states in details. In 
particular, we calculate mixed states of subsystems with finite rank symmetries and their 
infinite dimensional environment. We quantify their entanglement by determining their mutual 
information (entanglement entropy) and quantum negativity. We demonstrate the emergence of 
secondary parameters related to $\suinf$ symmetry and relative dynamics of subsystem. They 
include two continuous parameters, namely a quantity presenting relative area or size of 
diffeo-surfaces of $\suinf$ representations by subsystems, and a time variable. In addition, 
states may depend on two discrete parameters characterizing the orthogonal function basis used 
for construction of the generators of $\suinfa$ algebra. Mixed states of subsystems depend on 
these parameters and two continuous parameters characterizing $\suinf$ representations. We show 
that these states are similar to fields in QFT's with gauge symmetry, for example those of 
the Standard Model (SM). Thus, \sqgr~relates the origin of the observed gauge symmetries of 
fundamental interactions to QGR. On the other hand, quantum mixed nature of these states 
and their dependence on discrete quantum numbers, unrelated to their local symmetries, 
deviate from pure state of the SM particles. They reflect quantum origin of the 4 continuous 
parameters, which their effective values is shown in~\cite{houriqmsymmgr,hourisqgr} to be the 
perceived classical spacetime. These features might be testable in future experiments seeking 
signatures of quantum gravity.

\item{\bf Purification: } Quantum origin of the 4D continuous parameter space of subsystems 
is made transparent by purification of their mixed states. We demonstrate that states of 
subsystems satisfy conditions for perfect purification according to~\cite{qmpurifcond}. Moreover, 
using effective classical metric mentioned above, we show its relationship with the quantum 
state of environment seen by subsystems. In turn, we show that in accordance with results 
of~\cite{qmrefperspect} the mixed state of environment is perspective dependent and related to 
the internal symmetries of the subsystem. This is a remarkable result, because in general 
relativity, the relation between spacetime geometry and matter is established through dynamics, 
that is the Einstein-Hilbert action and Einstein equation. By contrast, in \sqgr~it arises 
through introduction of a quantum clock and before introducing an action.

\item{\bf Clarification of relationship between 4 continuous parameters of subsystems and their 
$\suinf$ symmetry: } Association of 4D continuous parameters to $\suinf$ symmetry of subsystems 
seems confusing, because representations of $\suinf$ depend on two continuous parameters. We show 
that purified states are composed of two entangled components, one representing a finite rank 
symmetry $G$ and the other $\suinf$. Using the fact that sine algebra represents 
$\suinfa + \mathcal{U}(1)$, we demonstrate that in presence of multiple representations of 
$\suinf$, their individual $U(1)$ symmetry breaks (see diagrams (\ref{singlesuinfdiag}) and 
(\ref{multisuinfdiag})), because one 
can define a morphism between their algebras - in other words between their associated 2D 
diffeo-surfaces. A consequence of this morphism is that relative area of these compact surfaces 
becomes an observable. Moreover, with introduction of a quantum clock and time parameter, quantum 
states of subsystems will depend on 4 continuous parameters, which are indistinguishable and 
quantum states are in their coherent superposition. Additionally, arbitrariness of references and 
clocks dictates that observables must be independent of reparameterization and diffeomorphism of 
the parameter space. Therefore, we conclude that 2D diffeo-surfaces of $\suinf$ representations 
are subspaces of the 4D parameter space (see Fig. \ref{fig:diffeo}). 

\item {\bf Equivalence of $\suinf$ gauge transformation and diffeomorphism of the parameter space 
of subsystems: } We provide an analytical description of the relationship between $\suinf$ 
symmetry of the Hilbert space and diffeomorphism of the space of its continuous parameters. 
We discuss both the case of the Universe as a whole, where there is only one representation 
of $\suinf$, and the case of the ensemble of subsystems representing $\suinf$ and their relative 
dynamics, where diffeo-surfaces of $\suinf$ representations are 2D subspaces of a 4D space. 
We show that diffeomorphism of this parameter space can be neutralized by a $\suinf$ gauge 
transformation. Therefore, its geometry is irrelevant for the Lagrangian functional of the model. 
This demonstration reinforce the proof of Proposition 2 of~\cite{hourisqgr}.
\end{description}

Additionally, we demonstrate the following general properties of $\suinf$ symmetry group and 
criteria for considering a quantum system as being composite. They were necessary for obtaining 
the results summarized in the previous paragraph. 

\begin{description}
\item{\bf Multiple limits of $\suinf$: } We show that the degeneracy of 
$\mathcal{SU}(N \rightarrow \infty)$ algebra demonstrated in~\cite{suninfrep} does not have 
impact on the construction and properties of \sqgr~parameter space. Although the sine 
algebras (\ref {suinftorustriang}) for $k \neq 2\pi/N$ where $N \in \mathbb{Z}$ are not 
homomorphic to each others or to $k = 2\pi/N$, they all have the same limiting algebra 
when $k \rightarrow 0$, namely (\ref{suinfatorus}) which is homomorphic to the algebra 
of area preserving diffeomorphism $ADiff (D_2)$ of 2D compact Riemannian surfaces. In the 
framework of \sqgr~it is this limit that we call $\suinfa$. Thus, there is no ambiguity in 
the definition of parameters characterizing representations of $\suinf$ in the model. 

\item{\bf Criteria for compositeness: } We show that two sets of conditions used in the 
literature as indicator of compositeness of a quantum system, namely factorization of density 
matrix or state vector~\cite{prodstate} and factorization of the algebra of 
observables~\cite{sysdiv} are equivalent. 
\end{description}

\section{Multiple limits of $\mathcal{SU}(N \rightarrow \infty)$ and their consequence for 
\sqgr}  \label{sec:suinfsqgr}
The group $\suinf$, its representations and algebra, and their properties are briefly reviewed 
in Appendix \ref{app:suinfrev}. Variables and symbols related to these entities 
are defined there.

A particularity of $SU(N \rightarrow \infty)$ is its non-uniqueness~\cite{suninfrep}. There 
are various ways to see this. For instance, if $N \rightarrow \infty$ and 
$N' \rightarrow \infty$ are prime with respect to each other, or more generally for irrational 
$k$'s and integer vectors $(\mathbf{m}, \mathbf{n})$ in (\ref{suinftorustriang}) - the case 
considered by~\cite{suinftriang0,adifftorussuinf,suninfrep} - structure coefficients of the 
algebras would be always different, and thereby they would not be homomorphic to each others. 
In~\cite{suninfrep} it is proved that for any $k \in \mathbb{C}$, the algebra 
(\ref{suinftorustriangint}) is homomorphic to $\suinfa + \mathcal{U}(1)$. Moreover, it is 
shown that for $k \in \mathbb{R}$, algebras with different values of $k$ are not pairwise 
homomorphic. Consequently, it seems that $SU(N \rightarrow \infty)$ and its algebra have 
uncountable many degeneracies. Additionally, for irrational $k$ such that 
$1/k \rightarrow \infty$ these algebras do not have a shift symmetry similar to 
(\ref{shiftsymm}) and cannot be homomorphic to $ADiff$ of any orientable Riemann surface. 
We remind that the proof of $\suinf \cong ADiff (D_2)$\footnote{In this work we use the 
symbol $\cong$ for homomorphism, because the symbol $\simeq$ used for this purpose in 
mathematics literature is used for numerical approximation in physics literature.} 
in~\cite{suninfsurfaceanomal,suninfrep0} is limited to the case of rational $k$ (or 
equivalently countable $N$ in (\ref{suinftorustriangint})). Furthermore, in Appendix 
\ref{app:sinalgebra}~ we demonstrate that the description of generators $\hK_{\mathbf{m}}$ of 
(\ref{suinftorustriang}) as functionals of $\mathbf{x}$ and $\partial_{\mathbf{x}}$ proposed 
in~\cite{suinftriang0} generates this algebra only approximately, and for $k \neq 0$ no other 
functional of these variables can generate the algebra exactly. This finding confirms 
the conclusion of~\cite{suninfrep}, asserting that some of (\ref{suinftorustriang}) algebras 
are not homomorphic to $ADiff(D_2)$.

In what concerns the application of $\suinf$ in string and matrix (M-)theories as candidate 
quantum gravity, the absence of its homomorphism with $ADiff (D_2)$ can be considered 
as a serious problem. Indeed, the motivation for investigating the relationship between 
$\suinf$ and $ADiff(D_2)$ in the 1980's and 1990's was using $N \times N$ matrices as 
symplectic approximation of strings worldsheets or more generally 
membranes~\cite{suninfsurfaceanomal,suninfsimplect}. In this context homomorphism of $\suinf$ 
and $ADiff(D_2)$ is crucial, because quantum gravitational interactions are associated to 
deformation of $D_2$ surfaces and vis-versa. Thus, homomorphism between geometrical and 
algebraic operations is crucial.

By contrast, $\suinf$ symmetry of the Hilbert space in \sqgr~arises because of the assumption 
of infinite number of mutually commuting observables. Although the homomorphism of $\suinf$ 
representations with $ADiff(D_2)$ is also important, specially for providing two 
continuous parameters, it is a subsidiary property. As discussed in details in~\cite{hourisqgr} 
and in the following sections here, these parameters along with two other continuous 
variables, namely area or size/distance and time, characterize $\suinf$ and dynamics related 
aspects of the quantum states of subsystems/particles. On the other hand, it is easily seen 
that $\suinfa$ algebra always depends on two variables, irrespective of the degeneracy of 
algebra (\ref{suinftorustriangint}) and whether it is homomorphic to $ADiff(D_2)$ or not. 
Indeed, $SU(N), \forall N$ is a $N$-dim representation of $SU(2)$. As the group manifold of 
$SU(2)$ is the sphere $S^{(2)}$, its generators in any representation, including $\suinf$, can 
be described as functions of angular coordinates of sphere, see~\cite{suninfhoppthesis} for 
details\footnote{It is easy to see that even without using association with $ADiff(D_2)$, 
algebras (\ref{suinftorustriangint}) and (\ref{suinftorustriang}) also depend on two 
approximately continuous variables. Indeed, generators $\hK_{\mathbf{m}}$ with 
$\sqrt{2\pi/N} \mathbf{m} \rightarrow (\infty, \infty)$ are dominant by their number. As 
fractional numbers are dense in $\mathbb{R}$, vectors $\sqrt{2\pi/N} \mathbf{m}$ are 
approximately continuous 2-vectors.}. Therefore, diffeomorphism of $\suinf$ and $ADiff(D_2)$ is 
less crucial for \sqgr.

In any case, it is important to remind that both (\ref{suinftorustriang}) and 
(\ref{suinftorustriangint}) algebras are homomorphic to $\suinfa$ only for 
$1/k \rightarrow \infty$. The expression (\ref{kgen}) of the generators of 
(\ref{suinftorustriang}) also becomes exact in this limit, which corresponds to the algebra 
(\ref{suinfatorus}) and homomorphic to $ADiff(D_2)$. Therefore, although for $k \neq 0$
(\ref{suinftorustriang}), (\ref{suinftorustriangint}), and (\ref{kgen}) present different 
algebras, they all converge to a unique limit, namely the algebra (\ref{suinfatorus}), which is 
homomorphic to Poisson algebra, and as demonstrated in Appendix \ref{app:suinfrev}, to 
$ADiff(D_2)$. Thus, from now on and in the context of \sqgr~model by $\suinfa$ we always mean 
this unique limit.

\section{Order from randomness: Emergence of subsystems}  \label{sec:subsysstate}
In~\cite{houriqmsymmgr, hourisqgr} we showed that \sqgr~Universe as a whole is static and 
trivial, in the sense that different states can be transformed to each others by application of 
$\suinf$ members, under which according to the axioms of the model the physics is invariant. 
Only when the Universe is divided to approximately isolated subsystems, structures and 
relative dynamics arise from their interactions and entanglement. In this section we study in 
more details quantum state of a subsystem and its relationship with the rest of the Universe, 
formed by the ensemble of other subsystems.

Consider a quantum system with a $N$-dimensional Hilbert space representing $SU(N)$ group. 
Assuming $N \gg 1$, the Cartan decomposition of $SU(N)$~\cite{cartandecomp} can be used to 
describe the symmetry, and thereby the Hilbert space, as tensor product of smaller rank 
Lie groups:
\be
\{|\psi_1 \rangle\} \times \{|\psi_2 \rangle \} \cdots \times \{|\psi_n\rangle \}  
\label{nbodybasis}
\ee
where the $n_i$-dimensional set $\{|\psi_i \rangle \}$ is a basis for the $i^{th}$ component 
of the Cartan decomposition. Notice that (\ref{nbodybasis}) has the same structure as the 
Hilbert space of an n-body system. In general, $N$, the number of components $n$, and 
dimensions $n_i$ can be infinite or even innumerable. In the case of \sqgr~ $N = n = \infty$. 

The Cartan decomposition of $\suinf$ group is reviewed in Appendix A of~\cite{hourisqgr}. 
It is shown that it can be decomposed to tensor product of infinite number of $SU(K)$, i.e 
$\suinf \cong \bigotimes^\infty SU(K)$ for any $K$, including infinity. More generally, 
$\suinf$ can be written as tensor product of any finite rank compact Lie group 
$G$~\cite{suninfrep0}. Thus, there is a basis similar to (\ref{nbodybasis}) for the Hilbert 
space of the Universe $\hm_U$ with $n_1 . n_2 . \cdots . n_{i \rightarrow \infty}$ components, 
corresponding to the decomposition $\suinf = G^1 \times G^2 \times \cdots$, where $n_i$ is 
dimension of representation $\{|\psi_i (k_i) \rangle \}$ for the group $G^i$. The $d_i$ 
dimensional set $\{k_i\}$ indicates space of parameters characterizing representation of $G^i$. 
With this decomposition the state of the Universe $|\Psi_U \rangle$ can be expanded as:
\be
|\Psi_U \rangle = \sum_{\{k_i = 1, \cdots, d_i \},~ i = 1, \cdots n\rightarrow \infty} 
\am (k_1 k_2 \ldots k_n, \cdots) ~ |\psi_{k_1}\rangle \times |\psi_{k_2}\rangle \times \cdots 
\times |\psi_{k_n} \rangle \times \cdots   \label{nbodystate}
\ee
Notice that we have simplified the notation by using $k_i$ as an index. Thus, 
$|\psi_{k_i}\rangle \equiv|\psi_i (k_i)\rangle$.

Description of a basis as a tensor product does not necessarily mean that the corresponding 
quantum system consists of subsystems~\cite{sysdiv}. In~\cite{houriqmsymmgr,hourisqgr} we 
used algebraic criteria introduced in~\cite{sysdiv} and random quantum fluctuations to explain 
how the concepts of locality and approximate division to subsystems arise in the Universe. 
An equivalent but more explicit criterion for definition of subsystems using a tensor product 
basis, as in (\ref{nbodystate}), is the factorization of amplitudes $A_{k_1 k_2 \cdots k_n}$. 
States for which these amplitudes can be decomposed to factors, depending only on a limited 
number of parameters that characterize $\otimes_i^{n \rightarrow \infty} |\psi_{k_i}\rangle$, are 
called {\it separable} or {\it product states}~\cite{prodstate}. They present partial 
localization in the Hilbert space. 

The factors $A_{k_1 k_2 \cdots k_n}$ depending on a single $k_i$ expand subspaces with no 
entanglement. By contrast, factors depending on $n' \rightarrow \infty$ parameters expand 
inseparable and fully nonlocal\footnote{Unless locality in spacetime is 
explicitly mentioned, here {\it locality} means in the Hilbert space, according to the 
definition given in Appendix \ref {app:ampfactor}.} subspaces of $\hm_U$. In Appendix 
\ref{app:ampfactor}~ we demonstrate that factorization of state vector or density matrix is 
equivalent to the algebraic criteria defined in~\cite{sysdiv} for the compositeness of a 
quantum system. 

\subsection{Decomposition to finite rank symmetries}  \label{sec:onesubsys}
In \sqgr~when dealing with the whole Universe it is not useful to consider a tensor 
decomposition like (\ref{nbodystate}) for its state $|\Psi_U\rangle$, because it 
does not survive a $\suinf$ transformation, which preserves physical observables. Indeed, 
in~\cite{hourisqgr,houriqmsymmgr} we demonstrated that due to this global symmetry 
$|\Psi_U\rangle$ is trivial - the vacuum - except for the dependence of $\suinf$ representation 
on the topology of its diffeo-surface. Nonetheless, states like $|\Psi_U\rangle$ can have 
very different components. Specifically, in~\cite{hourisqgr,houriqmsymmgr} we showed that 
quantum fluctuations, that is random application of operators $\hO \in \bm[\hm_U]$ on 
$|\Psi_U\rangle$, induce clustering of components. In particular, a completely coherent state 
in an arbitrary basis changes to a less coherent state with more clustered and unbalanced 
components. In such state, subsets of elements with very different amplitudes from 
their neighbours can be approximately considered as subsystems. Hence, the state 
$|\Psi_U\rangle$ can be approximately factorized to a tensor product of subspaces of 
$\hm_U$ where each factor represents a smaller rank symmetry.

Consider one of the factors in (\ref{nbodystate}) and assume that it represents a finite rank 
symmetry $G$. The ensemble of remaining factors is still infinite dimensional and represents 
$\suinf$~\cite{suninfrep0}. To prove this property we use decomposition of $\suinf$ to a 
finite rank $SU(K)$, i.e. $\suinf \cong \bigotimes^\infty SU(K), K \in \mathbb{Z}^++2$. 
For any finite rank compact Lie group $G$ one can find $K \in  \mathbb{Z}^++2$, such that 
$SU(K') \subset G \subset SU(K), ~ K' < K$, and the following relations are satisfied:
\be
\suinf \cong \bigotimes^\infty SU(K') \subseteq G \times \bigotimes^\infty SU(K') \subseteq 
G \times \bigotimes^\infty SU(K) \subseteq SU(K) \times \bigotimes^\infty SU(K) \cong \suinf  
\label{gsuinf}
\ee
Moreover, this chain shows that for any finite rank $G$:
\be
G \times \suinf \cong \suinf \label{gsuinfcong}
\ee
The difference between $\suinf$'s in the two extremes of (\ref{gsuinf}) is similar to a line 
without any specific point on it - analogous to the left extreme of (\ref{gsuinf}) - and a 
line with a finite number of selected points - analogous to the right extreme of (\ref{gsuinf}). 
In both cases the set of points constituting the line are the same. But, in the second case 
some points are interpreted, employed or perceived differently. A detailed demonstration of 
(\ref{gsuinfcong}) is provided in Appendix \ref{app:gsuingcongex}.

We call the basis that expands states representing $G$ symmetry $\{|\psi_G (k_G) \rangle\}$, 
where $k_G$ collectively represents parameters of the representation of $G$ by this basis. For 
finite rank Lie groups the number of independent $k_G$'s, that is the dimension $d_G$ of 
parameter space $\{k_G\}$ is finite. For instance, for $G= SU(2)$, 
$k_G = (l', m'), ~ l' \in \mathbb{Z}^+, ~ -l' \leqslant m \leqslant l'$. For a fixed $l'$, 
corresponding to a super-selected representation of $SU(2)$, the dimension $d_G = 2l' +1$. 

Following the tensor decomposition (\ref{gsuinf}), the state of the Universe can be expanded as: 
\be
|\Psi_U\rangle = \sum_{\{k_G\}, (\eta, \zeta, \cdots)\}} A (k_G; \eta, \zeta, \cdots) ~ 
|\psi_G(k_G) \rangle \times |\psi_\infty(\eta, \zeta; \cdots) \rangle  \label{gustate}
\ee
and the corresponding density matrix:
\bea
&& \hvarrho_U = \sum_{\substack {\{k_G, k'_G\} \\ (\eta, \zeta, \eta', \zeta', \cdots)\}}} 
A (k_G; \eta, \zeta, \cdots)  A^*(k'_G, \eta', \zeta', \cdots) ~ \hrho_G (k_G , k'_G) \times 
\hrho_\infty (\eta, \zeta, \eta', \zeta', \cdots) \label{gudensity} \\ 
&& \hrho_G (k_G , k'_G) \equiv |\psi_G (k_G) \rangle \langle \psi_G (k'_G)|, \quad 
\hrho_\infty (\eta, \zeta, \eta', \zeta', \cdots) \equiv |\psi_\infty (\eta, \zeta, \cdots) 
\rangle \langle \psi_\infty (\eta', \zeta', \cdots)|  \label{gudensitydef}  \\
&& \sum_{\substack {\{k_G\} \\ \{(\eta, \zeta, \cdots)\}}} 
|A (k_G; \eta, \zeta, \cdots)|^2 = 1  \label {amptracecond} 
\eea
where states $|\psi_\infty(\eta, \zeta; \cdots) \rangle$ constitute a basis for $\suinf$ factor 
of $\hm_U$ and continuous parameters $(\eta, \zeta)$ characterize generators of $\suinf$, see 
Appendix \ref{app:suinfrev}~ for details. Extension dots indicate secondary parameters defined 
in the following subsection. When they are irrelevant or do not contribute to the discussion 
we replace them with extension dots $''\cdots''$. Notice that we have used an arbitrary basis 
for $G$ and $\suinf$ factors. Consequently, in general off-diagonal elements of the density 
matrix are not zero. Moreover, despite the tensor product structure of the basis in 
(\ref{gustate}) and (\ref{gudensity}), amplitudes $A (k_G; \eta, \zeta, \cdots)$ are not 
exactly factorizable, because a global $\suinf$ transformation changes the tensor product 
decomposition of $|\Psi_U\rangle$ and $\hvarrho_U$. Therefore, isolation of a subsystem is 
always an approximation.

\subsubsection{Secondary parameters of $\suinf$ representations of subsystems} \label{sec:suinfbasis}
Secondary parameters in the basis $|\psi_\infty(\eta, \zeta; \cdots) \rangle$ are those that 
do not affect a single representation of $\suinf$, but can be involved in characterization 
of $\suinf$ represented by Hilbert spaces of multiple quantum systems/subsystems interacting 
with each others. One of these parameter $r$ indicates the relative area or a size scale for 
diffeo-surfaces of $\suinf$ representations in the model with respect to that of a reference. 
As discussed in Appendix \ref{app:suinfrev}, the algebra $ADiff(D_2)$ is homomorphic to 
$\suinfa + \mathcal{U}(1)$, where $\mathcal{U}(1)$ presents scaling of the diffeo-surface. 
This symmetry is irrelevant - unobservable - for the state of a single quantum system 
representing $\suinf$ irreducibly, such as $\hvarrho_U$. On the other hand, area or size 
scale of diffeo-surfaces of multiple (sub)systems representing $\suinf$ becomes a relative 
observable. This is the case when the Universe is divided to subsystems each representing 
$\suinf$ symmetry. 

Another secondary parameters is time $t$ characterizing the quantum state of a clock subsystem. 
In~\cite{hourisqgr,houriqmsymmgr} we showed that in this way subsystems acquire a relative 
dynamics \`a la Page \& Wootters~\cite{qmtimepage,qmtimepage0} or equivalent 
approaches~\cite{qmtimedef}. As time measurement may be POVM rather than projective, in 
\sqgr~quantum states of subsystems are in general in coherent superposition of 4 continuous 
parameters $x \equiv (\eta, \zeta, r, t)$, where $(\eta, \zeta)$ are parameters of $\suinf$ 
representation. Assuming compact ranges for the last two parameters, i.e. using the fact that 
$\mathbb{R} \cong U(1)$, or non-compact values for $\eta, \zeta$, the 4 parameters become 
indistinguishable components of a 4D vector 
$x \equiv (x^i,~i=0, \cdots, 3) \in S_4 (\mathbb{R}^4)$, of a compact or non-compact parameter 
space that we call $\Xi$. Logically, observables of subsystems should be invariant under 
reparameterization of $\Xi$. Moreover, Proposition 2 of~\cite{hourisqgr} demonstrates that 
the geometry of $\Xi$ is irrelevant for observables and can be modified by a $\suinf$ 
transformation under which the physics is invariant. This property will be further discussed 
in Sec. \ref{sec:gaugeequiv}.

As generators of $ADiff(D_2) \cong \suinf$ act on continuous functions of $(\eta, \zeta)$, 
they are usually expanded with respect to a set of orthogonal functions as basis. These 
functions may in turn depend on a set of usually discrete parameters that we collectively 
call $\mathbf{\ell}$. For example, for a sphere basis $\boldsymbol{\ell}$ corresponds to the 
spherical harmonic modes $(l,m)$, see appendices of~\cite{houriqmsymmgr} for a brief review. 
For a torus basis, as discussed in Appendix \ref{app:suinfrev}, 
$\boldsymbol{\ell} = \mathbf{m}$, where $\mathbf{m}$ is an integer 2D vector. We emphasize 
that $\boldsymbol{\ell}$ is not an independent parameter of the model, because as it is shown 
in~\cite{houriqmsymmgr}, the algebra $\suinfa$ and its generators can be formally expressed 
with respect to two continuous parameters only. 

\subsection{Quantum states of subsystems}  \label{sec:gstate}
To extract from the factorized $\hvarrho_U$ in (\ref{gudensity}) a state that represents only 
the group $G$, we trace out its $\suinf$ representing factor:
\be
\hvarrho_G \equiv \tr_\infty \hvarrho_U = \sum_{\substack{\{k_G, k'_G\}, \\ \{y\}}} 
A_G (k_G; y) A^*_G (k'_G, y) ~ \hrho_G (k_G , k'_G), \quad \quad y \equiv (\eta, \zeta, \cdots) 
\label{gdensity}
\ee
The amplitude $A_G$ differs from $A$ in (\ref{gudensity}) by a normalization factor, which may 
in general depends on $(\eta, \zeta, \cdots)$. We call the Hilbert space generated by 
$\hvarrho_G$ density matrices $\hm_G$. It represents the symmetry group $G$, because the basis 
$\hrho_G (k_G , k'_G)$ represents this symmetry. Nonetheless, amplitudes 
$A_G(k_G; \eta, \zeta, \cdots) $ and their complex conjugate $A^*_G$ depend on parameters 
$(\eta, \zeta, \cdots)$, which are not related to the $G$ symmetry. An observer ignoring their 
quantum origin interprets them as being related to a {\it classical} environment. On the other 
hand, application of any operator $\hO \in \bm[\hm_U] \cong \suinf$ changes $\hvarrho_U$, and 
consequently the dependence of $A_G (k_G; \eta, \zeta, \cdots)$ on the {\it environment} 
parameters $(\eta, \zeta, \cdots)$. Therefore, $\hvarrho_G$ presents the mixed quantum state 
of an open subsystem. To show explicitly the mixedness of $\hvarrho_G$ it is enough to 
demonstrate that $\hvarrho_G^2 \neq \hvarrho_G$. It straightforward to show that:
\bea
\hvarrho_G^2 &=& \sum_{\{\{y\}, \{\{y'\}} B(y, y') \sum_{\{k_G\}, \{k'_G\}} A_G (k_G; y) 
A^*_G (k'_G; y') ~ \hrho_G (k_G , k'_G)  \label{gudensity2} \\
B(y, y') & \equiv & \sum_{\{k''_G\}} ~ A^*_G (k''_G; y) A_G (k''_G; y')  \label{bigbdef}
\eea
Comparison of (\ref{gudensity2}) and (\ref{gdensity}) clearly shows that 
$\hvarrho_G^2 \neq \hvarrho_G$, except in the limit where amplitudes $A_G$ do not depend on 
the value of $y$. Thus, $\hvarrho_G$ is in general a mixed state and the subsystem representing 
$G$ is entangled to the rest of the Universe. 

In a similar manner ${\hrho_G (k_G)}$ factor in $\hvarrho_U$ can be traced out to find the 
contribution of $\suinf$ symmetry in $\hvarrho_U$ after separating a subsystem with $G$ 
symmetry:
\be
\hvarrho_\infty \equiv \tr_G ~ \hvarrho_U = \sum_{\substack{\{k_G\}, \\ 
\{(\eta, \zeta, \cdots)\}, \{(\eta', \zeta', \cdots)\}}} A_\infty (k_G; \eta, \zeta, \cdots) 
A^*_\infty (k_G; \eta', \zeta', \cdots) ~ \hrho_\infty (\eta, \zeta ; \eta', \zeta', \cdots) 
\label{infdensity}
\ee
We call the Hilbert space generated by $\hvarrho_\infty$ states $\hm_\infty$. Using analogous 
expressions to (\ref{gudensity2} - \ref{bigbdef}) for $\hvarrho_\infty$, it is straightforward 
to show that $\hvarrho_\infty$ is also a mixed state. Moreover, it depends on 
{\it external parameters} $k_G$, which is not related to $\suinf$ symmetry. In Appendix 
\ref{app:entangle}~ we calculate several entanglement measures for $\hvarrho_G$ and 
$\hvarrho_\infty$.

\subsubsection{Significance of entanglement}  \label{sec:entanglesig}
For various reasons the above results are interesting. They show that although $\hm_U$, and 
thereby $\hvarrho_U$, are infinite dimensional and after tracing out a finite number of its 
components the resulting state $\hvarrho_\infty$ is still infinite dimensional, it is no longer 
pure. Indeed, application of any $\hO \in G \subset \suinf$ to $\hvarrho_U$ 
changes the dependence of $A_\infty (k_G; \eta, \zeta, \cdots)$ on $k_G$, which is the 
{\it environment} parameter for $\hvarrho_\infty$. Therefore, the {\it environment} itself is 
an open subsystem of the Universe. The only {\it closed} - and thereby static - quantum 
system is the Universe itself. This observation is an explicit demonstration of the 
Proposition 1 in~\cite{hourisqgr}, and confirms the entanglement of every subsystem with the 
rest of the Universe. 

Another interesting aspect of $\hvarrho_\infty$ is its dependence on the parameters of the 
traced symmetry $G$, which as we have discussed in~\cite{hourisqgr} (and further in the 
following sections) can be considered as an approximately isolated subsystem - a matter 
field. In~\cite{houriqmsymmgr,hourisqgr} we related the variation of $\hvarrho_\infty$ to an 
affine parameter and an effective metric, which we identified as the metric of the perceived 
classical spacetime:
\be
ds^2 \equiv \Lambda ds_{WY}^2 = \Lambda ~ \tr (\sqrt{\delta\hvarrho_\infty} 
\sqrt{\delta\hvarrho_\infty}^\dagger) \equiv g_{\mu\nu}(y) dy^\mu dy^\nu  \label{metric}
\ee
The constant $\Lambda$ is a dimensionful scaling constant and $ds_{WY}^2$ is the separation 
between $\hvarrho_\infty$ and $\hvarrho_\infty + \delta\hvarrho_\infty$ according to the 
Wigner-Yanase skew information~\cite{wigneryanaseqminfo,qmspeedlimgen}. Using Mandelstam-Tamm 
inequality~\cite{qmspeed} - also called quantum speed limit - we proved that the effective 
metric $g_{\mu\nu}$ is Lorentzian and has negative signature. Thus, it is reasonable to interpret 
it as the perceived classical spacetime, that we call $\Xi_{sp}$ to distinguish it from the 
parameter space $\Xi$. 

The dependence of $\hvarrho_\infty$ and its variation $\delta\hvarrho_\infty$ on the matter 
parameter $k_G$ means that in \sqgr~relationship between the effective metric $g_{\mu\nu}$ 
and state of matter emerges at constructional level. By contrast, in Einstein gravity metric 
tensor and affine parameter are related to matter and its state dynamically, that is through 
the Einstein-Hilbert action and Einstein equation. Of course, the single equation (\ref{metric}) 
is not enough for a full determination of $g_{\mu\nu}$ tensor and dynamics is needed for this 
purpose.

\subsection{Emergence of gauge symmetries}  \label{sec:ymgauge} 
In quantum mechanics and quantum information inaccessible or unobservable degrees of 
freedom are usually traced out. Hence, we can interpret separation/tagging of a finite 
dimensional subspace representing group $G$ from $\hm_U$ and tracing out its infinite 
dimensional complement subspace as isolation of a {\it subsystem} due to the 
{\it inaccessibility} of its {\it environment}. 

According to (\ref{gsuinf}), it is possible to factorize infinite number of $G$ factors from 
$\suinf$. Each factor can be considered as one realization of a {\it subsystem} representing 
$G$. Moreover, $\hm_G$ may be a reducible representation of $G$. In this case $\hvarrho_G$ can 
be interpreted as many-body quantum state of an ensemble of subsystems each representing $G$ 
irreducibly. In both cases subsystems are not completely isolated and interact with each 
others through application of operators $\hO_G \in \bm[\hm_G]$. They also interact with each 
others and with other types of subsystems with different internal symmetries through their 
shared $\suinf$ symmetry/gravity.

In (\ref{gdensity}) operators $\hO_G$ only affect $\hrho_G$. Therefore, they can be 
performed independently for every value of the {\it environment} parameters 
$(x; \boldsymbol{\ell})$. By rearranging summations in (\ref{gdensity}) the density 
matrix $\hvarrho_G$ can be written as:
\bea
\hvarrho_G & = & \sum_{\substack{\{k_G\}, \{k'_G\} \\ \{(x; \boldsymbol{\ell})\})\}}} 
\hat{\upvarrho}_G (k_G, k'_G; x; \boldsymbol{\ell}) \label{rhogrdecomp} \\
\hat{\upvarrho}_G (k_G, k'_G; x; \boldsymbol{\ell}) & \equiv & A_G (k_G; x; \boldsymbol{\ell})
A^*_G (k'_G; x; \boldsymbol{\ell}) ~ \hrho_G (k_G , k'_G) \label{gdensityym}
\eea
Operators $\hat{\upvarrho}_G$ have the structure of local quantum fields in a 4D Yang-Mills QFT 
with group $G$ as their gauge symmetry, in the sense that application of  the members of $G$ 
symmetry projects them to the space generated by $\hrho_G$\footnote{We remind that the notion 
of density matrix in quantum mechanics and notion of quantum field in QFT are related. Appendix 
\ref{app:fielddens}~ review of their relationship.}. We emphasize on the locality of such operations, 
because they can be performed independently for each set of parameters $(x; \boldsymbol{\ell})$ 
a subsystem. In~\cite{hourisqgrcomp,hourisqgr} we showed that in order to have a dynamics 
invariant under reparameterization of $\Xi$, the operator $\hat{\upvarrho}_G$ must transform 
similar to a spin-1 field (see also Sec. \ref{sec:gaugeprop}). This completes the demonstration 
of the analogy between $\hat{\upvarrho}_G (k_G, k'_G; x; \boldsymbol{\ell})$ and Yang-Mills 
QFT's. However, there are several differences between 
$\hat{\upvarrho}_G (k_G, k'_G; x; \boldsymbol{\ell})$ and fields in models without quantum 
gravity. In \sqgr~the locality of finite rank $G$ symmetry arises at construction level. This 
is in contrast to QFT's in which the dynamics - the Lagrangian - is designed such that it be 
invariant under local application of symmetry operators. Additionally, in QFT's there is no 
discrete parameters analogous to $\boldsymbol{\ell}$, we call {\it mode quantum numbers} of 
$\suinf$ symmetry - quantum gravity.

\subsection{Purified states of subsystems}  \label{sec:subsuinf}
In~\cite{hourisqgr,houriqmsymmgr} we used the global $\suinf$ symmetry and how it affects 
subsystems to demonstrate that full symmetry of subsystems is $G \times \suinf$. In this section 
we use purification of the mixed state $\hvarrho_G$ to obtain this symmetry from a quantum 
point of view rather than algebraic properties of the model. 

Although tracing of $\suinf$ related factors in (\ref{gudensity}) demonstrates how a local 
Yang-Mills structure associated to a finite rank symmetry arises for subsystems, it erases the 
quantum origin of $\suinf$ and dynamics related parameters $x;\boldsymbol{\ell}$. Nonetheless, 
with the prior knowledge that the origin of these parameters in $\hvarrho_G$ is the global 
$\suinf$ symmetry of the Universe, we can extend $\hm_G$ by a Hilbert space representing 
$\suinf$. The new Hilbert space $\hm_{G_\infty}$ represents $G \times \suinf$ and its states 
can be expanded as:
\be
|\Psi_{G_\infty}\rangle \equiv \sum_{\{k_G\}; (\eta, \zeta, \cdots)\}} 
A_{G_\infty} (k_G; \eta, \zeta, \cdots) ~ |\psi_G (k_G) \rangle \times 
|\psi_\infty (\eta, \zeta, \cdots) \rangle  \label{gpurified}
\ee
where $|\psi_\infty (\eta, \zeta, \cdots)$ is the same as in (\ref{gustate}). Notice that 
$|\Psi_{G_\infty}\rangle$ can be also considered as purification of $\hvarrho_\infty$. Although 
the state $|\Psi_{G_\infty}\rangle$ looks like the state of the Universe $|\Psi_U\rangle$ in 
(\ref{gustate}), they do not present exactly the same quantum system for following reasons.

In (\ref{gpurified}) the group $\suinf$ acts only on the $|\psi_\infty (\eta, \zeta, \cdots) \rangle$ 
factor of the basis. Thus, in comparison with the state $|\Psi_U\rangle$ in (\ref{gustate}), 
or equivalently its density matrix $\hvarrho_U$ in (\ref{gudensity}), the global $\suinf$ symmetry 
is broken. The difference between $|\Psi_{G_\infty}\rangle$ and $|\Psi_U\rangle$ becomes 
more transparent if we express $\suinf$ as $SU(N \rightarrow \infty)$. Indeed, in 
$|\Psi_{G_\infty}\rangle$ the $\suinf$ symmetry of the appended Hilbert space should be read as 
$SU((N - d_G) \rightarrow \infty)$. In addition, the basis of the appended subspace 
$|\psi_\infty (\eta, \zeta, \cdots) \rangle$ is not necessarily the same as that used for obtaining 
$\hvarrho_G$. Therefore, amplitudes $A_{G_\infty} (k_G; \eta, \zeta, \cdots)$ are arbitrary up to 
the unitarity constraint:
\be
\sum_{\substack {\{k_G\} \\ \{(\eta, \zeta, \cdots)\}}} |A_{G_\infty} (k_G; \eta, \zeta, \cdots)|^2 = 1 
\label {rhoginfunitconstraint}
\ee
and {\it faithfulness} constraint~\cite{qmpurifcond}, that is tracing over extended subspace 
must recover $\hvarrho_G$:
\be
\tr_\infty \hvarrho_{G_\infty} = \hvarrho_G, \quad \quad \hvarrho_{G_\infty} \equiv |\Psi_{G_\infty}\rangle 
\langle \Psi_{G_\infty}|  \label{rhoginfconstraint}
\ee
These conditions are not sufficient to uniquely relate amplitudes $A_{G_\infty}$'s to $A_G$'s. 
This is not a surprise as the degeneracy of purification is well known through Sch\"odinger-HJW 
theorem~\cite{qmpurif,qmpurif0,qmpurif1} asserting that any mixed state can be purified to 
many unitarily equivalent pure states. This is a consequence of the arbitrariness of both the 
{\it auxiliary} space used for the extension of the Hilbert space, and the basis chosen for 
this subspace. Although here we have chosen a fixed auxiliary Hilbert space based on our prior 
knowledge about how $\hvarrho_G$ was obtained, as mentioned earlier, the basis remains 
arbitrary. It is well known that the lack of information about the basis - the reference 
frame - leads to decoherence~\cite{qmfreframe}. Specifically, because the basis 
$|\psi_\infty (\eta, \zeta, \cdots) \rangle$ is orthogonal to the $G$-related factor and can be 
chosen or modified arbitrarily, the global $SU(N \rightarrow \infty)$ coherence symmetry is 
broken to $G \times SU((N - d_G) \rightarrow \infty)$. Nonetheless, if we treat the two 
subspaces representing $G$ and $SU((N - d_G) \rightarrow \infty)$ as subsystems, they remain 
entangled because the constraint (\ref{rhoginfconstraint}) ensures a perfect and faithful 
purification. Details of conditions for faithful purification and their satisfaction by 
$\hvarrho_G$ and $\hvarrho_\infty$ is discussed in Appendix \ref{app:purifcond}.

\subsubsection{Physical interpretation of purified states}  \label{sec:purifinterpret}
Using linearity of the map $\Uplambda$ and the last equality in (\ref{purifmap}), which is the 
condition for faithfulness, it is straightforward to show that either $\Uplambda$ maps all 
$\hvarrho_G$ to the same pure density matrix, or it cannot be a universal purifier for an 
ensemble of closely related mixed density matrices~\cite{qmpurifcond}\footnote{Here by 
closeness we mean density matrices that can be possible states of a given quantum system or 
subsystem. If states $\hrho_1$ and $\hrho_2$ are in the set of possible states for a system, 
their sum can be also a plausible state. The proof of Theorem 1 in~\cite{qmpurifcond} uses 
this superposition property.}. Thus, when it is a universal purifier, its outcome would be a 
pure quantum state independent of details of $\hvarrho_G$. For $\hvarrho_G$ in a given 
representation of $G$ symmetry, amplitudes $A_{G_\infty} (k_G; \eta, \zeta, \cdots)$ would 
be independent of $A_G (k_G; \eta, \zeta, \cdots)$ up to a unitary transformation of 
$\hrho_G \times \hrho_\infty$ basis. Physically, this means that when subsystems with the same 
internal symmetry are approximately disentangled from the rest of the Universe, they see the 
same environment, and therefore are indistinguishable. This conclusion is also consistent 
with the global equivalence of all $\hvarrho_U$ states up to a physically irrelevant unitary 
transformation. By contrast, subsystems with different internal symmetries or different 
representations of the same symmetry see the rest of the Universe differently. Consequently, 
their purifying map $\Uplambda$ would be different, because their projections into the 
extended Hilbert space would be different. 

In quantum information and quantum technologies purification of the states of quantum systems 
is necessary for compensating noise and decoherence induced by interaction with an 
uncontrollable environment. Here, the purification of states $\hvarrho_G$ and $\hvarrho_\infty$ 
has rather a conceptual purpose. It shows that in a Universe with infinite number of mutually 
commuting observables, pure quantum states can be associated to subsystems and their 
environment, at least intermittently. Indeed, the random action of global operators $\hO \in \suinf$, 
which can be considered to be part of the environment, affects both. On the other hand, similarity of 
the purified subsystem and environment states demonstrates that they are not really separable, 
and irrespective of what happens to them and the way their states are presented, their 
entanglement is retained.

\subsection{Emergence of area/size parameter}  \label{sec:ensemblestate}

In Sec. \ref{sec:subsuinf}~ we explicitly showed that in \sqgr~subsystems of the Universe 
represent $G \times \suinf$. This means that the Hilbert space $\hm_{G_\infty}$ of each subsystem 
is a representation of $\suinf$. In~\cite{hourisqgr,houriqmsymmgr} we argued that in presence 
of multiple subsystems representing $\suinf$, the area of their diffeo-surfaces becomes a 
relative observable and should be added to the list of parameters characterizing their states. 
Here we investigate this property further.

The Poisson algebra (\ref{suinfal}) of $ADiff(D_2)$, which is homomorphic to the algebra 
$\suinfa$~\cite{suninfhoppthesis,suinftriang0} is invariant under scaling. Thus, (\ref{suinfal}) 
and associated algebras, that is the Poisson algebra of spherical harmonics in sphere 
basis~\cite{suninfhoppthesis}, algebras (\ref{suinftorustriang}) and (\ref{suinftorustriangint}) 
in torus basis, and their limit (\ref{suinfatorus}) are homomorphic to the Lie algebra 
$\suinfa + \mathcal{U}(1)$ and represent the symmetry group 
$\suinf \times \mathbb{R} \cong \suinf \times U(1) = U(\infty)$. The scaling symmetry 
$\mathbb{R} \cong U(1)$ reflects the irrelevance of the area of compact diffeo-surfaces $D_2$ 
in the limit where these algebras are homomorphic to $ADiff(D_2)$. 

It is easy to prove that $\otimes^n \suinf \cong \suinf,~ \forall n$ see e.g. appendices 
of~\cite{hourisqgr}. This relation can be also expressed as:
\be
\otimes^n \suinf \cong \bigcup_{i=1}^n ADiff(D^{(i)}_2) \cong ADiff(D_2), \quad \quad 
D_2 \equiv  \bigcup_{i=1}^n  D^{(i)}_2   \label{suinfdiffeo}
\ee
In isolation the area of each diffeo-surface $D^{(i)}_2$ is arbitrary and its variation does not 
affect the homomorphism of $ADiff(D_2^{(i)}$ with $\suinf$. But, once diffeo-surfaces are stuck 
together, only the area of their ensemble $D_2$ can be arbitrarily scaled. Thus, scaling 
diffeo-surface of one subsystem must be necessarily compensated by scaling of others. The 
algebra becomes:
\be
\otimes^n (\suinfa + \mathcal{U}(1)) \rightarrow \otimes^n \suinfa + \mathcal{U}(1) \cong \suinfa + \mathcal{U}(1)  
\label{suinfadiffeo}
\ee
The consequence of these properties is the dependence of quantum states of subsystems to an 
additional continuous parameter $r > 0$ that indicates relative area (or its square root) 
of compact diffeo-surfaces of subsystems. Notice that $r = 0$ is equivalent to trivial 
representation of $\suinf$ and is excluded by axioms of the model. The following diagrams 
summarize the relationship of $\suinf$ and $ADiff$ for single and multiple representations:

\bigskip

\begin{center}
\begin{tabular}{c}
Single subsystem  \\
\begin{tikzcd} 
  \suinf  \dar[swap, "\cong" ] \longrightarrow \suinfa \rar["\text{area irrel.}" ] 
  &  \suinfa + \um (1) \dar["\cong" ] \\
  ADiff(D_2)~ \rar[swap, "\text{area irrel.}"]
  & ~ ADiff(D_2)\times U(1)
\end{tikzcd} \\
\end{tabular}
\be {\label{singlesuinfdiag}} \ee
\end{center}

\begin{center}
\begin{tabular}{c}
Multiple subsystems \\
\begin{tikzcd}[column sep=huge]
\suinf \times \cdots \times \suinf \cong \suinf \longrightarrow \suinfa 
\rar["\text{area irrel.}" ] \dar[swap, "\text{area irrel.}"]  
  & \suinfa + \um (1)  \dar["\text{area irrel.}"] \\
(ADiff(D^{(1)}_2) \times U(1)) \times \cdots \times (ADiff (D^{(n)}_2) \times 
U(1)) \arrow[r, "\text{symm. break}" swap, "\text{area preserv.}"] 
& ADiff(D_2) \times U(1) \\
\end{tikzcd} 
\end{tabular}
\end{center}
\vspace{-0.5cm}
\be \label{multisuinfdiag}  \ee


\subsubsection{Consistency of the two perceptions of the quantum Universe} \label{sec:univconsist}
Once the quantum state of the Universe is self-clustered to approximately isolated subsystems 
and a clock parameter $t$ is chosen, the purified state of the ensemble of subsystems 
$\hvarrho_{U_s}$ can be written as: 
\bea
\hvarrho_{U_s} & = & \bigotimes_{i=1}^\infty \hvarrho_{G^{(i)}_\infty} =  
\bigotimes_{i=1}^\infty \sum_{\substack {\{k_{G^{(i)}}, k'_{G^{(i)}}\} \\ (x_i; \boldsymbol{\ell}_i;x'_i; 
\boldsymbol{\ell}'_i)\}}} 
A (k_{G^{(i)}}; x_i; \boldsymbol{\ell}_i) A^*(k'_{G^{(i)}}; x'_i; \boldsymbol{\ell}'_i) 
\biggl (\hrho_{G^{(i)}} (k_{G^{(i)}} , k'_{G^{(i)}}) \times \hrho_\infty (x_i; \boldsymbol{\ell}; 
x'_i; \boldsymbol{\ell}'_i) \biggr )  \nonumber \\
& = & \prod_i \biggl (\sum_{\substack {\{k_{G^{(i)}}, k'_{G^{(i)}}\} \\ (x_i; \boldsymbol{\ell}; x'_i; \boldsymbol{\ell}'_i)\}}} 
A (k_{G^{(i)}}; x_i; \boldsymbol{\ell}_i) A^*(k'_{G^{(i)}}, x'_i; \boldsymbol{\ell}'_i) \biggr ) 
\bigotimes_{i=1}^\infty \hrho_{G^{(i)}} (k_{G^{(i)}} , k'_{G^{(i)}}) \times 
\bigotimes_{i=1}^\infty \hrho_\infty (x_i; \boldsymbol{\ell}; x'_i; \boldsymbol{\ell}'_i) \nonumber \\
\label{ustatemultiple}
\eea
Here we have assumed that each subsystem $i$ may represent a different internal finite rank 
symmetry $G^{(i)}$. However, from cosmological and laboratory observations we know that many 
subsystems/particles share the same internal symmetry. Therefore, many of $G^{(i)}$'s and their 
representations in (\ref{ustatemultiple}) are the same. In addition, the number of subsystems 
does not need to be countable.

In appendices of~\cite{hourisqgr} we demonstrated that tensor products 
$\otimes_{n \rightarrow \infty} G \cong \suinf$ for any $G$, including $G = \suinf$ 
(see also~\cite{suninfrep0}). Hence, both factors in the tensor product in the last equality 
of (\ref{ustatemultiple}) are density matrices representing $\suinf$. Even if the time parameter 
$t$ is fixed\footnote{Assuming a fixed value for the time parameter $t$, rather than integrating 
it, means a projective measurement of the clock observable associated to $t$. Nonetheless, the 
state of the clock for this outcome should be included in $\hvarrho_{U_s}$, because the clock may 
have other observables commuting with that used as time. For instance, if $t$ corresponds to the 
phase of an oscillating spin of an atom, its uncorrelated (commuting) degrees of freedom to 
spin, such as kinetic energy would be in general in coherent superposition and full quantum 
state of the clock would be similar to (\ref{gpurified}).}, despite apparent dependence of 
$\hvarrho_{U_s}$ on time it belongs to the same Hilbert space $\hm_U$ as the state $\hvarrho_U$ 
in (\ref{gudensity}), namely the quantum state of the entire Universe without breaking it to 
subsystems. Specifically, due to $\otimes^\infty_{i=1} G_i \cong \suinf$, the state 
$\hvarrho_{U_s} (t) \in \hm_{U(t)} \subset \hm_U$, and can be transformed to any other state of 
$\hm_U$ by application of $\suinf$ group, which according to axioms of \sqgr~should not affect 
observables. The state of the Universe can be evolved only if the clock is approximately 
isolated from other subsystems. In this case, as explained in Sec. (\ref{sec:gstate}), it will 
remain entangled to the rest of the Universe and can evolve by both internal - 
self-interaction - and interaction with other subsystems through both shared internal symmetries 
and $\suinf$ - gravity. 

In conclusion, considering the Universe as the ensemble of its infinite number of subsystems 
is not in contradiction with treating it as a single quantum system with infinite number of 
mutually commuting observables. The difference of perception is on whether concentration or 
access is on the local features of the Universe's state or its global - topological - properties. 
In what concerns our present observational capabilities, we can only detect its local aspects.

\section {Dynamics over a parameter space}  \label{sec:dyn}
Quantum state of the ensemble of subsystems $\hvarrho_{U_s} (t)$ calculated in 
(\ref{ustatemultiple}) includes all information about them. However, it is not suitable for 
applications, because it does not show how subsystems interact with each others and evolve 
with respect to the clock. In quantum mechanics and QFT the evolution is usually formulated 
through definition of an action functional or a Hamiltonian in non-relativistic cases, which 
are usually taken or inspired from classical limit of the model. However, \sqgr~is built as a 
quantum theory and it does not have a classical formulation, or even a true classical limit, 
see footnote \ref{foot:classiclim}~ and~\cite{hourisqgr}. Therefore, dynamics must be constructed 
solely based on the axioms of the model and properties concluded from them.

In~\cite{hourisqgr} we introduced dynamics by constructing a Lagrangian functional over 
parameter space $\Xi$. It consists of symmetry invariant traces of $\suinf$ generators, and 
those of internal symmetry $G$ when the Universe is divided to subsystems. After imposing 
invariance under reparameterization of $\Xi$, we showed that at lowest order in traces it has 
the form of a Yang-Mills theory for both symmetries. Traces of multiple generators of Lie groups 
are functions of their rank and structure coefficients. Moreover, in the Lagrangian functional 
such terms are proportional to higher orders of coupling constant. Therefore, trace terms 
with more than two generators can be considered as perturbative corrections to lowest order 
effective Lagrangian. Consequently, methodology and techniques of QFT's such as path integral 
and Feynman diagrams are applicable to \sqgr. Importantly, as Yang-Mills models are known to be 
renormalizable \sqgr~does not have any issue in this regard.

In this section after a brief reminding of the \sqgr~action we elaborate some of details which 
were not addressed in the previous works.

\subsection{\sqgr~as a Yang-Mills QFT on the parameter space of subsystems} \label{sec:gaugeprop}
\begin{figure}
\begin{center}
\includegraphics[width=14cm]{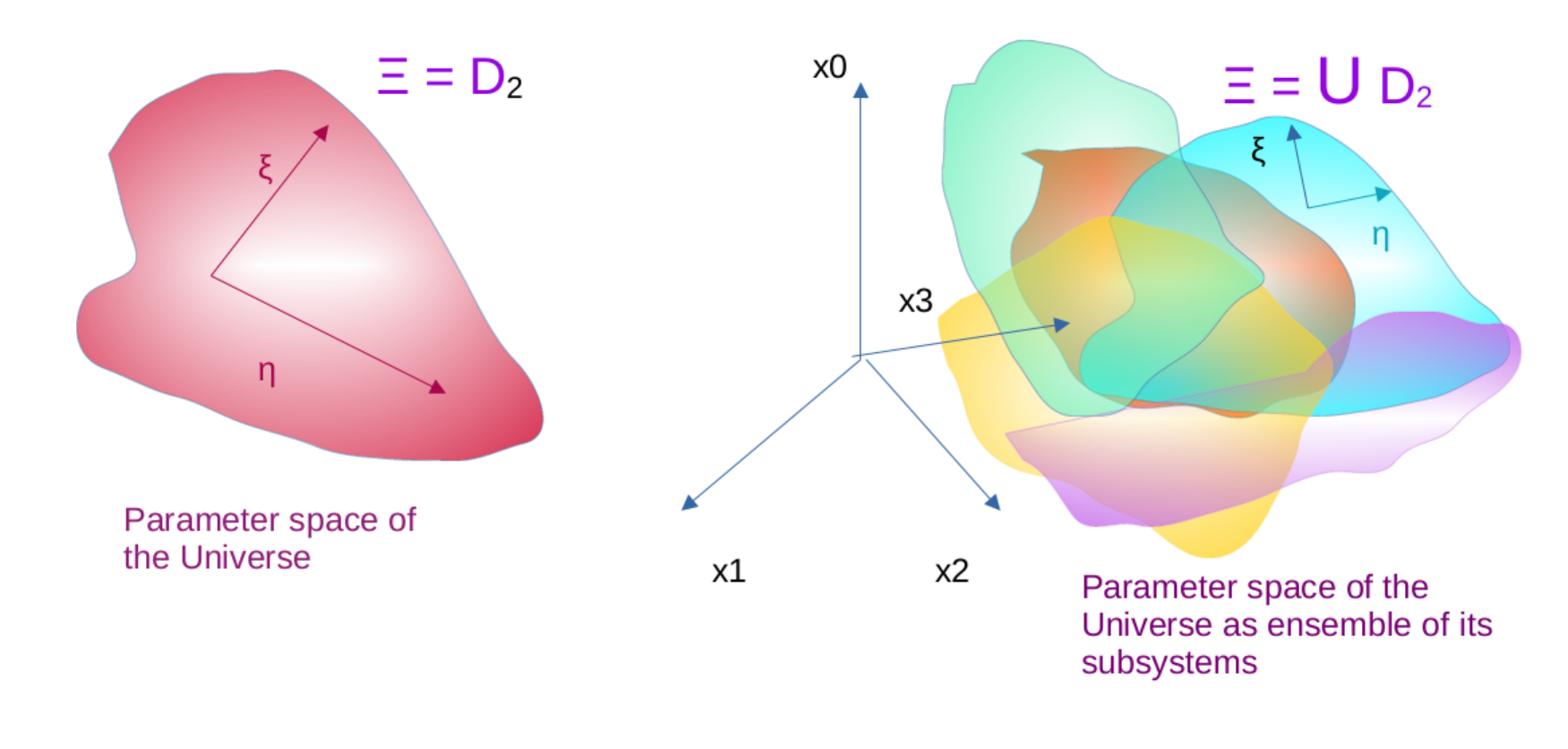} 
\end{center}
\vspace{-1cm}
\caption{Schematic presentation of parameter spaces $\Xi$ of $\suinf$ symmetry. Left: A single 
2D compact surface corresponding to diffeo-surface of $\suinf$ symmetry of the Universe. Its 
area is arbitrary and irrelevant. Axes $(\eta, \zeta)$ are local coordinates. Right: The 4D 
parameter space generated by infinite number of 2D diffeo-surfaces of subsystems and time 
parameter of a quantum clock. Axes $x^i,~i=0, \cdots, 3$ are coordinates at each point 
$x \in \Xi$ and axes $\eta (x), \zeta (x)$ are induced coordinates on one of the diffeo-surfaces 
passing through $x$. \label{fig:diffeo}}
\end{figure}

The effective Lagrangian of \sqgr~is analogous to the effective Lagrangian of QFT's, with 
the difference that it is defined on the parameter space $\Xi$, rather than an external 
spacetime. Reparameterization and symmetry invariance lead, at lowest order, to a Yang-Mills 
type Lagrangian functional on $\Xi$. Specifically, for the whole Universe, that is when the 
Universe is treated as a single isolated quantum system, the lowest order effective Lagrangian 
$\lm_U$ is a 2D Yang-Mills QFT on the diffeo-surface $\Xi = D_2$ (see Fig. \ref{fig:diffeo}~ 
for a schematic depiction):
\bea
\lm_U & = & \int d^2\Omega ~ \biggl [ ~\frac{1}{2} ~ \tr (F^{\mu\nu} F_{\mu\nu}) + 
\frac {1}{2} \tr (\sD \hvarrho_U) \biggr ], \quad \quad 
d^2\Omega \equiv d^2x \sqrt{|\upeta^{(2)}|} \label{lagrange2d} \\
F_{\mu\nu} & \equiv & F_{\mu\nu}^a \hL^a \equiv [D_\mu, D_\nu], \quad 
D_\mu = (\partial_\mu - \Gamma_\mu) + \sum_a i \lambda A_\mu^a (x) \hL^a (x), 
\label{yminvardef}
\eea
where $\hL^a (x)$ are generators of $\suinf$ symmetry, $\Gamma_\mu$ is a suitable connection 
for $\Xi$, and reparameterization invariant derivative $\sD$ depends on how $\hvarrho_U$ is 
transformed under diffeomorphism of $\Xi$. See the next paragraph for 
definition of $\sD$. Notice that both the gauge field $A_\mu^a$ and generators $\hL^a$ depend 
on the coordinates of the parameter space - the diffeo-surface. For instance, for the sphere 
basis~\cite{suninfhoppthesis} (reviewed in~\cite{houriqmsymmgr}) $x \equiv (\theta, \phi)$ are 
angular coordinates on $\Xi$, index $a = (l,m)$, and $\hL_{lm} (\theta, \phi)$ are generators 
of $\suinf$. In torus basis reviewed in Appendix \ref{app:suinfrev}, $x$ is the 2D Cartesian 
coordinates of a point on $\Xi$, index $a = \mathbf{m}$, and $\hL_{\mathbf{m}}$ is defined in 
(\ref{torusbasis}). In $\suinf$ Yang-Mills on a background spacetime of any dimension, 
in addition to the {\it internal coordinates} $x$, the gauge field depends on the spacetime, 
see Appendix \ref{app:ymspacetime}~ for a brief review of these models. The Lagrangian $\lm_U$ 
is clearly static, because it does not include any time parameter. Moreover, it has been shown 
that it is topological~\cite{houriqmsymmgr,hourisqgr} and its value does not depend on the 
2D metric $\upeta_{\mu\nu}, ~ \mu, \nu = 1, 2$. 

When the Universe is considered as ensemble of its subsystems representing $G \times \suinf$, 
as explained in Sec. \ref{sec:subsysstate}, their purified states $\hvarrho_{G_\infty}$ depends, 
in addition to coordinates of the diffeo-surface, on an area/size parameter and on time. The 
symmetry and reparameterization invariant Lagrangian $\lm_{U_s}$ is defined on the 4D parameter 
space $\Xi$ (depicted in Fig. \ref{fig:diffeo}), and has the following expression:
\bea
\lm_{U_s} & = & \int d^4 x \sqrt{|\upeta|} ~ \biggl [\tr (F^{\mu\nu} F_{\mu\nu}) + 
\frac{1}{4} \tr (G^{\mu\nu} G_{\mu\nu}) + \frac {1}{2} \sum_s \tr (\sD \hvarrho_{G_\infty}) \biggr ] 
\label{yminvarsub} \\
F_{\mu\nu} (x) & \equiv & [D_\mu, D_\nu], \quad D_\mu = (\partial_\mu - \Gamma_\mu) - 
 i \lambda A_\mu, \quad A_\mu \equiv \sum_a A_\mu^a (x; \eta, \zeta) 
\hL^a (\eta, \zeta) \label{yminvardefsuinf} \\
G_{\mu\nu} (x) & \equiv & [D'_\mu, D'_\nu], \quad D'_\mu = 
(\partial_\mu - \Gamma_\mu - i \lambda A_\mu) \mathbbm{1}_G - 
i \lambda_G B_\mu, \quad B_\mu \equiv \sum_b B_\mu^b (x) \hT^a \label{yminvardefg}
\eea
Definitions of quantities in this Lagrangian are as follows. The symmetric tensor $\upeta_{\mu\nu}$ 
is the metric of the parameter space $\Xi$. It should not be confused with the metric 
$g_{\mu\nu}$ of the emergent classical spacetime $\Xi_{sp}$. According to the Proposition 2 
of~\cite{hourisqgr} $\upeta_{\mu\nu}$ is arbitrary, because geometry of $\Xi$ is not a physical 
observable. We further discuss this property in Sec. \ref{sec:gaugeequiv}. The first and second 
terms in (\ref{yminvarsub}) are the Lagrangian density for the $\suinf$ and internal symmetry 
$G$ gauge fields $A_\mu^{lm}$ and $B_\mu^a$, respectively. The $\suinf$ generators $\hL^{lm}$ and 
the range of $(l,m)$ are defined in Appendix \ref{app:suinfrev}. $\Gamma_\mu$ is the geometric 
connection of the parameter space $\Xi$. Operators $T^a$ are generators of the internal 
symmetry $G$ of subsystems. Their number is determined by the range of index $a$, which must 
be finite because  the rank of $G$ is finite. The density matrix $\hrho_s$ is the quantum 
state of a subsystem. The covariant operator $\sD$ is a differential operator and its exact 
definition depends on the representation of the symmetries of $\Xi$ by the states of 
subsystems. For instance, for spinors $\sD \equiv \gamma^0 \gamma^i e_i^\mu (\partial_\mu - \Gamma_\mu) \mathbbm{1} - \sum_{lm} i \lambda_s A_\mu^{lm} \hL^{lm} - \sum_a i \lambda_G B_\mu^a \hT^a$, 
where $\gamma^\mu, \mu = 0, \cdots, 3$ are Dirac matrices and $e_i^\mu, i=0, \cdots, 3$ are 
tetrads, see~\cite{hourisqgr} for more details. The reason for the presence of $\suinf$ 
field in (\ref{yminvardefg}) and $\sD$ is the fact that all subsystems represent $\suinf$.
In Sec. \ref{sec:gaugeequiv}~ we demonstrate 
that geometry connection terms in $D'_\mu$, $\sD$ and field equations can be neutralized 
by a $\suinf$ gauge transformation. Therefore, as mentioned above the Lagrangian 
(\ref{yminvarsub}) does not depend on $\upeta_{\mu\nu}$. Finally, notice that the interaction 
of $G$ gauge field (amplitude) $B_\mu$ with $\suinf$ is implicit, because it depends on the 
parameters $x \in \Xi$ of the $\suinf$ symmetry of subsystems.

Aside of including terms for Yang-Mills fields representing $G$, the main difference between 
the $\suinf$ gauge field in $\lm_{U_s}$ and $\lm_U$ is its dependence on additional parameters. 
In $\lm_U$ the gauge field depends only on two continuous parameters characterizing $\suinfa$ 
algebra. By contrast, $A_\mu^a$ in $\lm_{U_s}$ depends on 4D vectors $x \in \Xi$ and on 
$(\eta, \zeta) \in D_2$, where $D_2$ is the diffeo-surface of $\suinf$ symmetry it represents. 
Although the $\suinf$ gauge field (\ref{gaugedef}) defined on a background spacetime has 
apparently a similar structure, there is an important difference. In contrast to coordinates of 
so called {\it internal space} in (\ref{gaugedef}), the induced coordinates $(\eta, \zeta)$ 
of the diffeo-surface $D_2$ are functions of $x$. As the Fig. \ref{fig:diffeo}~ shows, $D_2$ is 
immersed in $\Xi$, that is $D_2 \subset \Xi$. By contrast, in (\ref{gaugedef}) the internal 
space is independent of the background spacetime. It is the reason for using the name 
{\it diffeo-surface} rather than {\it internal space} for a compact 2D surface that its 
$ADiff$ group is homomorphic to $\suinf$. 

\subsubsection{Classical spacetime from quantum gravity action}  \label{sec:qgrmetric}
Perturbation theory applied to $\lm_{U_s}$ allows to calculate components of $\hvarrho_{G_\infty}$. 
Tracing out the contribution of $G$ symmetry parameters gives $\hvarrho_\infty$, which can be 
then used to determine the effective classical metric in (\ref{metric}). However, this procedure 
only determines $ds^2$ rather than the metric $g_{\mu\nu}$ of the emergent classical spacetime 
$\Xi_{sp}$. For determining components of $g_{\mu\nu}$ we have to use the classical limit of 
$\lm_{U_s}$ as explained in~\cite{hourisqgr}. The procedure consists of treating pure $\suinf$ 
term in (\ref{yminvarsub}) as the classical scalar curvature, other terms as {\it matter}, and 
identifying the metric of parameter space $\upeta_{\mu\nu}$ with the effective metric $g_{\mu\nu}$ 
defined in (\ref{metric}).However,the $\suinf$ gauge field $A_{\mu}$ must be included in the 
connection $\Gamma_\mu$, because in classical gravity, it is the connection that convoys 
gravitational interaction. The effective energy-momentum tensor should be determined from 
effective Lagrangian for matter part of (\ref{yminvarsub}), which can include quantum corrections 
in a curved spacetime, see e.g.~\cite{renormadiab0,renormadiab1,houriqmcond} for examples. Then, 
the Einstein equation obtained from applying variational principle to this approximation can be 
solved to calculate components of the effective metric. Of course, the Lagrangian $\lm_{U_s}$ can 
be used to calculate Green's functions and other measurables of both quantum matter and quantum 
gravity - the $\suinf$ gauge field - without having to consider the classical limit.

\subsection {Equivalence of diffeomorphism of parameter space and $\mathbf{\suinf}$ gauge transformation}  \label{sec:gaugeequiv}
The equivalence of reparameterization of the parameter space $\Xi$ and transformation of its 2D 
subspaces under the action of $\suinf$ group is necessary for proving that geometry of $\Xi$ 
is irrelevant for physics (Proposition 2 in~\cite{hourisqgr}). It is straightforward to verify 
that the connection form $\Gamma_\mu$ does not affect gauge invariant field intensity 
$F_{\mu\nu}$. But, it appears in its covariant derives as well as in the covariant derivative 
$\sD$ of the density matrix. In Einstein gravity metric, connection, and curvature tensors of 
the background spacetime are related to the energy-momentum tensor of matter through 
Einstein-Hilbert action and Einstein equation. By contrast, in \sqgr~geometry of parameter 
space $\Xi$ is arbitrary and should not affect observables. In~\cite{hourisqgr} we used algebraic 
arguments to prove this property. Here we demonstrate it analytically.

Equivalence of the diffeomorphism of $\Xi$ and $\suinf$ gauge transformation in the Lagrangian 
$\lm_U$ of the whole Universe is trivial, because $\Xi$ is the same as the diffeo-surface $D_2$. 
Thus, area preserving diffeomorphisms of $\Xi$ are simply $\suinf$ transformations, under which 
$\lm_U$ is invariant. To see this, consider a change of $\Gamma_\mu \rightarrow \Gamma'_\mu$ that 
preserves the area of $\Xi=D_2$. We remind that vectors in $\hm_U$ are 2D complex-valued 
functions\footnote{Nonetheless, in~\cite{hourisqgr} we showed that phase of the gauge fields is 
irrelevant and without loss of generality these fields can be considered to be real.}. 
Therefore, in analogy with $A_\mu$, components of $\Gamma_\mu$ can be considered as vectors of 
$\hm_U$. Using, for instance, the sphere basis for $\suinf$, both $\Gamma_\mu$ and $A_\mu$ can 
be expanded with respect to spherical harmonic functions $Y_{lm} (\theta, \phi)$. As 
$\hL_{lm} Y_{l'm'} = -i f ^{l"m"}_{lm,l'm'} Y_{l"m"}$ (for $\hbar = 1$) where $f ^{l"m"}_{lm,l'm'}$ 
are structure coefficients of $\suinf$~\cite{suninfhoppthesis}, a $\suinf$ gauge 
transformation of $A_\mu$ changes coefficients of its expansion with respect to 
$Y_{lm} (\theta, \phi)$. Thus, a gauge transformation such that:
\bea
(\Gamma'_\mu - \Gamma_\mu) &=& \Upomega (x) A_\mu (x) \Upomega^{-1} (x) -i \Upomega (x) 
\partial_\mu \Upomega^{-1} (x) - A_\mu (x) \label{connctdiff} \\
\Upomega (x) & \equiv & e^{i \int_{D_2} d^2 \Omega' ~ \epsilon_{lm} (\theta, \phi, \eta', \zeta') \hL^{lm} 
(\eta', \zeta')}  \label{gaugedef}
\eea
can compensate the change of $\Gamma_\mu$. A general diffeomorphism of any manifold $\mm$ can be 
decomposed to:
\be
Diff (\mm) \cong ADiff (\mm) \times \mathtt{\Lambda} (\mm), \quad \forall \mm 
\label{diffeodecomp}
\ee
where $\mathtt{\Lambda} (\mm)$ presents the operation of a global scaling of the manifold 
$\mm$. Such transformation changes $\lm_U \rightarrow \mathtt{\Lambda} \lm_U$. In other words, 
it changes the volume of $\Xi$ by rescaling its coordinates. But, due to the $U(1)$ symmetry 
discussed in Sec. \ref{sec:ensemblestate}, such a rescaling is not an observable.

Demonstration of the invariance of Lagrangian $\lm_{U_s}$ of subsystems and the corresponding 
field equations under diffeomorphism of their 4D parameter space $\Xi$ is more complicated, but 
follows the same line of reasoning. As discussed above, the 2D diffeo-surface $\Sigma$ 
characterizing $\suinf$ generators is immersed in $\Xi$, that is $\Sigma \subset \Xi$. Thus, the 
induced coordinates $(\eta,\zeta)$ of $\Sigma$ are functions of the 4D parameter space $\Xi$. 
To demonstrate the irrelevance of the geometry of $\Xi$ for physical observables, we show that 
gauge transformations involving all parameters can restore curvatures, geometrical connection, 
covariant derivatives of deformed or reparameterized $\Xi$\footnote{Riemann and Ricci curvatures 
depend on the derivatives of Levi-Civita connection. For this reason, calculating connection as 
a function of curvatures amount to solving a partial differential equation, which is not a 
trivial task. Therefore we take the inverse approach and prove that curvatures are restored by 
$\suinf$ gauge transformation.}. Specifically, Riemann and Ricci curvature tensors - hence the 
connection at a point $x$ of a Riemannian manifold - can be determined from sectional curvatures 
of 2D surfaces passing through that point~\cite{seccurvproof,seccurvproof0}.
There are exactly 6 such pairs of axes at each point $x$ of the 4D parameter space $\Xi$ leading 
to the following expression for sectional curvatures:  
\be
K(\Pi_{ij}(x)) \equiv K(\hat{x}^i,\hat{x}^j) = \frac{R_p(\hat{x}^i, \hat{x}^j, \hat{x}^i, 
\hat{x}^j)}{\langle \hat{x}^i, \hat{x}^i\rangle \langle \hat{x}^j, \hat{x}^j\rangle - 
\langle \hat{x}^i, \hat{x}^j\rangle^2}, \quad \quad {i,~j = 0, \cdots 3}, \quad i \neq j 
\label{sectionalcurv}
\ee
The 4D unit vectors $\hat{x}^i \in T\Xi (x)$, where $T\Xi (x)$ is the tangent space of $\Xi$ at 
$x \in \Xi$, and the plane $\Pi_{ij} (x)$ passes through vectors $\hat{x}^i$ and $\hat{x}^j$. 
As the 2D compact surfaces used for the definition of $\hL^a (\eta, \zeta)$ in 
(\ref{yminvardefsuinf}) are arbitrary, we can locally identify them with surfaces generated 
by a pair of unit axes $(\hat{x}^i,\hat{x}^j) \in T\Xi (x)$ at $x$. Moreover, as 
$\suinf^n \cong \suinf$, the union of diffeo-surfaces is again a diffeo-surface associated 
to a representation of $\suinf$. Thus, $\suinf$ gauge transformations, such as:
\bea
A_\mu (x; \eta, \zeta) & \rightarrow & A'_\mu (x; \eta, \zeta) = 
\Upomega (x; \eta, \zeta) A_\mu (x; \eta, \zeta) \Upomega^{-1} (x; \eta, \zeta) -i 
\Upomega (x; \eta, \zeta) \partial_\mu \Upomega^{-1} (x; \eta, \zeta) \nonumber \\ 
\label{gaugetrans4d} \\ 
\Upomega (x; \eta, \zeta) & \equiv & 
\exp(i\epsilon_{lm} (x; \eta, \zeta) \hL^{lm} (\eta, \zeta)), \quad x \in \Xi  
\label{gaugedef4d} 
\eea
can be performed using 
$(\eta, \zeta) \in \text{Span}(\hat{x}^i,\hat{x}^j),~i,j = 0, \cdots 3, i \neq j$, which means 
$(\eta, \zeta)$ are chosen such that the surface passing through them corresponds to $\Pi_{ij} (x)$ 
at $x$. Consequently, by properly choosing a set of $\suinf$ gauge transformation 
(\ref{gaugetrans4d}), Riemann and Ricci curvatures of the deformed or reparameterized $\Xi$ can 
be restored. Hence, even in the case of Lagrangian $\lm_{U_s}$ that depends on 4 parameters a 
$\suinf$ gauge transformation is homomorphic to an $ADiff$ of the parameter space and vis-versa. 
In other words, a change in the field space that preserves the Lagrangian $\lm_{U_s}$ is 
equivalent to a diffeomorphism of its parameter space and vis-versa. This demonstration 
finalizes analytic proof of of the Proposition 2 in~\cite{hourisqgr} and shows that the 
geometry of $\Xi$ is irrelevant for physical observables of subsystems. Thus, without loss of 
generality we can consider a fixed geometry for $\Xi$ in both the Universe as a whole, and 
when it is perceived through the ensemble of its subsystem. This choice is equivalent to fixing 
$\suinf$ gauge.

\section{Perspectives for test and applications of \sqgr} \label{sec:outline} 
Currently there is no observed evidence of quantum behaviour of gravity.  However, the 
existence of a quantum description for the universal interaction called gravity seems 
inevitable~\cite{qmgrinconsist,qmgrinconsist0,qmgrinconsist1}. Therefore, the absence of 
evidence is most probably due to the insufficiency of precision and resolution of current 
experiments. Nonetheless, with the huge progress of quantum technologies in the past few 
decades there is strong hope that in the near future tests of the quantumness of gravity and 
eventually discrimination between various QGR proposals become achievable. Indeed, there are 
already some progress in this direction. Specifically, cosmological and astro-particle 
observations put stringent constraint on the modification of dispersion relation of high energy 
photons due to the {\it micro-structure} of the spacetime~\cite{grestgrb090510a} predicted by 
some QGR models such as loop quantum gravity; gravitational waves constrain graviton 
mass~\cite{grbphotonmass} and Lorentz invariance violation~\cite {grlorentz} predicted by some 
string theories; large extra dimensions are constraint by various analyses and observations, 
for example~\cite{braneinfconst,houribraneqcd,branesncmbconst}.

In what concerns the test of \sqgr, as mentioned earlier, its most discriminative signature is
a spin-1 boson as mediator of QGR. However, observation of this attribute in laboratory 
experiments - usually based on the detection of a change in 
coherence~\cite{qgrtestdecoher,qgrtestdecoher0,qgrtestdecoher1} or 
entanglement~\cite{qgrtestentangle,qgrtestentangle0}\footnote{There are many proposals for test 
of QGR in the literature. The citations here are only examples and far from being exclusive.} 
of the state of a quantum system by gravitational interaction - is much harder. Meanwhile and 
until the limited capabilities of laboratory experiments become available, cosmological 
observations may be a faster road to detection of a QGR signal, in particular that of \sqgr. 
Here we outline a proposal for the detection of QGR signature using coherent astronomical sources.

\subsection{Difference in coherence of astronomical masers as a signature of QGR} \label{sec:maserlens}
Astronomical masers have short duration quantum coherence~\cite{masercoher} and can be used as 
a coherent source for testing QGR. The change of their polarization may be also useful for 
estimate spin of particles which led to their decoherence~\cite{qgrtestentangle0}. On their 
path from the place of their formation - usually in molecular clouds of galaxies at cosmological 
distances - to Earth, maser photons may be affected by gravitational field of another galaxy or 
galaxy cluster and their dark matter halos and being lensed~\cite{lensedmaser}. 

In a quantum view, the deflection of photons is the result of their interaction with gravitons. 
In general, such process changes the coherence of photons. Specifically, a quantum process 
(channel) affecting a quantum system can be described by a set of Krauss operators: 
$\hrho_f = \sum_i \hK_i \hrho_i \hK_i^\dagger$, where $\hrho_i$ and $\hrho_f$ are the initial and 
final states of the system, respectively, and Krauss operators $\hK_i$, satisfying 
$\sum_i \hK_i \hK_i^\dagger = \mathbbm{1}$ define the quantum process. Using the sum of 
off-diagonal components $C \equiv \sum_{i \neq j} \hrho_{ij}$ as a measure of the coherence of 
a state $\hrho$~\cite{qminfocohere}, it is evident that in general $C_f \neq C_i$. 

In the framework of \sqgr, the $\suinf$ gauge-invariant Krauss operators for the process of 
photon deflection by a gravitational field can be written as: 
$\hK_{\mu\nu}^a \equiv \nm F_{\mu\nu}^a \hL^a$, where $\nm$ is a normalization constant, 
$a \equiv (l,m)$ defined in Appendix \ref{app:suinfrev}~ is the $\suinf$ graviton {\it color}, 
and $F^a_{\mu\nu}$ is the spin-1 graviton field defined in (\ref{yminvardefsuinf}). Similarly, 
the $U(1)$ gauge invariant photon state can be expanded with respect to $G_{\mu\nu}$ defined 
in (\ref{yminvardefg}): $\hrho = \nm' \hG_{\mu\nu} \hG^{\mu\nu}$, where $\nm'$ is a normalization 
constant. A hat is added to emphasize that the field should be considered as an operator in 
$\bm[\hm_{G=U(1)]}$. 


Due to the deflection of maser's photons by the gravitational field of a galaxy or cluster 
playing the role of a gravitational lens, the optical paths of photons in different images are 
not the same. Consequently, the Krauss operators affecting them would be different. Thus, 
coherence of the maser emission in the images would not be the same. Therefore, detection of 
coherence difference in lensed masers - after taking into account other sources of coherence 
distortion - would be a signature of QGR. Another measure of the quantum effect of gravity on 
the lensed photons is entanglement fidelity~\cite{enranglfidel} (also called channel 
fidelity~\cite{channelfidel}) defined as 
$F_e \equiv \sum_i \tr(\hrho \hK_i) \tr(\hrho \hK^\dagger_i)$. This quantity measures how much a 
quantum channel changes the state of a signal. Assuming that before interacting with 
gravitational field of the lens the maser photons have the same single-photon state, the 
difference between channel fidelity of images would be due to the different Krauss operators 
of the channel (path) taken by photons in separate images\footnote{Photons of a maser beam are 
entangled by their phase. A priori the difference between entanglement entropy of images could 
be used as a complementary signature of QGR. However, different optical path of images, short 
coherence time of the maser photons, and different arrival time of the images means that their 
entanglement before getting lensed cannot be used.}. In contrast to the measurement of 
coherence that only depends on the quantum state of detected photons, channel fidelity depends 
also on the properties of the channel, in particular the spin of gravity mediator. But, it needs 
modeling of the lens and is more sensitive to its uncertainties. Nonetheless, a priori it should 
make possible to discriminate between different QGR models, in particular to test the prediction 
of a spin-1 intermediate boson in \sqgr.

This test of QGR has the advantage to be intrinsically multi-messenger. Incoherent radiation 
from the galaxy or cluster can be used to model the lens and it gravitational field, and various 
baryonic effects on the maser photons, which would be the main source of uncertainty for the 
observation of a tiny QGR signal. The large volume and depth of future radio surveys should be 
able to detect with better resolution more lensed galaxies with maser sources.

\subsection{Perspective for further investigations}  \label{sec:perspect}
Aside from seeking best methods for testing \sqgr, future works should investigate formulation 
of dynamics of subsystems as open quantum systems, because there is no truly isolated quantum 
system in the Universe. An open system approach would be particularly necessary for the design 
of experimental setups for the test of \sqgr~and other QGR models, for example the experiment 
proposed in~\cite{qgrtestopen}. Other topics to be studied, both for verifying the capability 
of \sqgr~to solve puzzles of QGR and for determining its predictions include: physics of quantum 
black holes, their evaporation, and puzzles of apparent information loss and singularity; 
QGR effects in the early Universe and inflation; nature of dark energy and whether it has a 
QGR origin; particle physics in the framework of \sqgr~- in other words how the Universe was 
fragmented to subsystems, emergence of local symmetries which have ultimately led to the 
Standard Model at low energies; and potential role of QGR in matter-antimatter symmetry breaking.

\section*{Acknowledgments}
The author acknowledges the support of the Institut Henri Poincar\'e UAR 839 CNRS-Sorbonne 
Universit\'e), and the LabEx CARMIN (ANR-10-LABX-59-01).

\section*{Funding statement}
This work did not have dedicated funding.

\section*{Conflicts of Interest statement}
This work does not have any known conflict of interest.

\section*{Data Availability Statement}
Results of this work can be freely used if the present work is cited.

\appendix
\section{A brief review of $\mathbf{\suinf}$ representations}  \label{app:suinfrev}
The symmetry group $\suinf$ and its relevance for physics were first noticed in the 1980's 
when J. Hoppe~\cite{suninfhoppthesis} demonstrated its homomorphism, in what is called 
{\it sphere basis}, with the group of area-preserving diffeomorphism $ADiff$ of 2D sphere 
$S^{(2)}$, see e.g. appendices of~\cite{houriqmsymmgr} for a brief review of its algebra. In 
this basis the algebra does not allow a central charge, and thereby is an infinite dimensional 
algebra independent of Virasoro algebra. The latter is another infinite dimensional algebra 
extensively studied and employed in physics since 1970's for construction and quantization of 
string or more generally membrane theories and random matrices or M-theories. Indeed, the main 
subject of interest for~\cite{suninfhoppthesis} was field theories on 2D (pseudo)-Riemannian 
surfaces, assumed to be immersed in multi-dimensional Lorentzian manifolds. Follow-up 
investigations of $\suinf$ were specially motivated by infinite dimensional matrices as 
symplectic approximation of 2D surfaces, $SU(N \rightarrow \infty)$ symmetry represented by such 
matrices, and their diffeomorphism as 2D gravity~\cite{suninfvirasoro,suninfvirasoro0,suninfsimplect,suinftriang,suinftriang0,suninftorus}. 

Consider the area integral of a 2D compact surface $D_2$ immersed in a $(1, d-1)$ dimensional 
Minkowski space $M$. The induced metric on $D_2$ can be described with respect to coordinates 
$X_\alpha,~\alpha = 0, \cdots, d-1$ of $M$~\cite{suninfhoppthesis,suninfvirasoro}. 
The coordinates $X_\alpha$ on $D_2$ are scalar functions of local coordinates $(\eta, \zeta)$ 
of $D_2$ and time. Thus, the area integral can be considered as action for dynamical variables 
$X_\alpha$ and their dynamics can be obtained by application of variational principle. 
The solution of field equations corresponds to minimizing or fixing the area of the membrane 
$D_2$, and it must be invariant under reparameterization of its coordinates. This implies 
that the Jacobian of transformation 
$(\eta, \zeta) \rightarrow (f (\eta, \zeta), g(\eta, \zeta))$ must be unity:
\be
\text{Det} \biggl (\frac{\partial (f, g)}{\partial (\eta, \zeta)} \biggr ) = \frac{\partial f}
{\partial \eta} \frac{\partial g}{\partial \zeta} - \frac{\partial g}{\partial \eta} 
\frac{\partial f}{\partial \zeta} \equiv \{f, g\} = 1 \label{poissonbr}
\ee
where $\{f, g\}$ is the Poisson bracket of functions $f$ and $g$. It can be 
shown~\cite{suninfhoppthesis,suninfvirasoro} that infinitesimal area preserving coordinate 
transformations, that is $ADiff(D_2)$ can be generated by operators having the following form:
\be
\hL_f = \frac{\partial f}{\partial \eta} \frac{\partial }{\partial \zeta} - 
\frac{\partial f}{\partial \zeta} \frac{\partial }{\partial \eta} \quad , 
\quad \hL_f ~ g = \{f,g\} \label{suinfgendef}
\ee
where $f$ is any $C^\infty$ scalar function on $D_2$. The Lie algebra of these operators is: 
\be
[\hL_f, \hL_g] = \hL_{\{f,g\}},  \label{suinfal}
\ee
The functions $f$ and $g$ can be in general complex valued. This is crucial when this algebra 
and its associated group is assumed to be the symmetry of the Hilbert space of a quantum 
system.

Functions on $D_2$ can be locally decomposed to any orthogonal function basis. However, area 
of a compact surface is a non-local property and topology of $D_2$ may dictate a globally 
preferred basis. For instance, for $D_2 = S^{(2)}$  a suitable basis for definition of 
generators $\hL_f$ in (\ref{suinfal}) is the set of spherical harmonic functions 
$Y_{lm} (\theta,\phi)$, where $\theta$ and $\phi$ are angular coordinates of $S^{(2)}$. Thus, 
$(\eta, \zeta) \equiv (\cos \theta, \phi)$ and generators are indexed by 
$l = 0, 1, \ldots, ~ -l \leqslant m \leqslant l$. In~\cite{suninfhoppthesis} it is demonstrated 
that the Lie algebra of generators $\hL_{lm} (\theta, \phi)$ is homomorphic to the algebra 
$\mathcal{SU} (N \rightarrow \infty)$. They have the expression (\ref{suinfgendef}) for 
$f = Y_{lm} (\theta,\phi)$. Properties of this representation of $\suinf$ and its application 
to the Hilbert space of the Universe and its subsystems in the framework of \sqgr~are studied 
in some details in~\cite{hourisqgr,houriqmsymmgr}.

Another interesting example is $D_2 = T^{(2)}$, a 2D torus. In this case, the globally valid 
basis has the form of a 2D harmonic oscillator~\cite{suninfvirasoro,suninfvirasoro0,suinftriang,suninftorus,suninfsurfaceanomal}:
\be
f (\eta, \zeta) \equiv \exp [i(m_1 \eta + m_2\zeta)] = \exp (i\mathbf{m}.\mathbf{x}), 
\quad \mathbf{m} \equiv (m_1,m_2), \quad \mathbf{x} \equiv (\eta, \zeta). \label{torusbasis}
\ee
and the algebra (\ref{suinfal}) becomes:
\bea
[\hL_{\mathbf{m}},\hL_{\mathbf{n}}] & = & (\mathbf{m} \times \mathbf{n})~\hL_{\mathbf{m+n}}  
\label{suinfatorus} \\
\hL_{\mathbf{m}} & = & -i \exp (i\mathbf{m}.\mathbf{x}) (\mathbf{m} \times \partial_{\mathbf{x}}), 
\quad \partial_{\mathbf{x}} \equiv (\partial/\partial \eta, \partial/\partial \zeta). 
\label{hlmdef}
\eea
The symbol $\times$ means vector product in 2D. For instance, 
$\mathbf{m} \times \mathbf{n} \equiv m_1 n_2 - m_2 n_1$. It is 
demonstrated~\cite{adifftorussuinf} that $ADiff(T^{(2)})$ is also homomorphic to $\suinf$. 
The difference between $SU(N \rightarrow \infty)$ related to $ADiff(S^{(2)})$ and that related 
$ADiff(T^{(2)})$ is in the way the number of roots in the $SU(N \rightarrow \infty)$ Dynkin 
diagram is extended to infinity~\cite{adifftorussuinf}. The group $ADiff(S^{(2)})$ is 
homomorphic to $\suinf_+$, corresponding to one side extension of Dynkin's diagram to infinity. 
By contrast, $ADiff(T^{(2)})$ is homomorphic to a $SU(N \rightarrow \infty)$ which its Dynkin's 
root diagram is extended to infinity from both sides. These results show that there are more 
than one $\suinf$ group and they may not be homomorphic to each others.

The algebra (\ref{suinfatorus}) can have a central charge without violating its homomorphism 
with $ADiff(T^{(2)})$~\cite{suninfvirasoro0}. This property demonstrates its relationship with 
Virasoro algebra, when the components of indices $\mathbf{m}$ and $\mathbf{n}$ are 
integer~\cite{suninfvirasoro,suninfvirasoro0}. Moreover, in~\cite{suinftriang} it is shown 
that this algebra is a special case of a more general class of algebras of the form:
\be
[\hK_\mathbf{m}, \hK_\mathbf{n}] = c(\mathbf{m},\mathbf{n}) \hK_{\mathbf{m+n}}  \label{viralgeblike}
\ee
The Jacobi identity of structure coefficients $c(\mathbf{m},\mathbf{n})$ is solved by 
$c(\mathbf{m},\mathbf{n}) \propto (m-n)$, corresponding to Virasoro algebra, and by 
$c(\mathbf{m},\mathbf{n}) \propto \sin (k \mathbf{m} \times \mathbf{n})$:
\be
[\hK_{\mathbf{m}},\hK_{\mathbf{n}}] = r \sin (k (\mathbf{m} \times \mathbf{n})) \hK_{\mathbf{m+n}}  
\label{suinftorustriang}
\ee
It is straightforward to see that (\ref{suinfatorus}) is a special case of 
(\ref{suinftorustriang}) in which $r=1/k$ and $k \rightarrow 0$. It is demonstrated that these 
algebras are homomorphic to $\suinf$~\cite{suinftriang0,suninfrep}. Here we review the proof 
for the case of 
$k = 2\pi/N, ~N \in \mathbb{Z}, ~ (m_1,m_2),~ (n_1,n_2) \in (\mathbb{Z}, \mathbb{Z})$:
\be
[\hK_{\mathbf{m}},\hK_{\mathbf{n}}] = (\frac{N}{2\pi}) \sin (\frac{2\pi}{N} (\mathbf{m} \times 
\mathbf{n})) \hK_{\mathbf{m+n}}  \label{suinftorustriangint}
\ee
There is a generalization of Pauli matrices for generators of $SU(N)$~\cite{sungen,sungen0}:
\bea
&& \cg \equiv \diag\{1, \omega, \omega^2, \cdots, \omega^{N-1} \}, \quad \quad 
\omega \equiv 2\pi i/ N, \\
&& \ch \equiv \begin {bmatrix} 0 & 1 &0 & \ldots \\ 0 & 0 & 1 & 0 \ldots \\  \vdots \\  
0 & 0 &0 & \ldots 1 \\ 1 & 0 & 0 & 0 \ldots \end{bmatrix} \\
&& \ch \cg = \omega \cg \ch, \quad \quad \cg^N = \ch^N = \mathbbm {1}.  \label{sunpauli}
\eea
In this basis unitary unimodular $N \times N$ matrices, that is unitary matrices with 
determinant $\pm 1$ representing $SU(N)$ group can be factorized to $\cg$ and $\ch$. In 
particular, matrices (operators) $\hK_\mathbf{m}$ satisfying the algebra 
(\ref{suinftorustriangint}) with integer or more generally fractional $N$ can be written as:
\be
\hK_\mathbf{m} = \omega^{m_1m_2} \cg^{m_1}\ch^{m_2}  \label{kintop}
\ee
Using (\ref{sunpauli}) it is straightforward to show that these operators satisfy 
(\ref{suinftorustriangint}). Hence, this algebra is isomorphic to $\suinf$ when 
$N \rightarrow \infty$, or equivalently $k = 2\pi/N \rightarrow 0$. In this case the algebra 
(\ref{suinftorustriangint}) asymptotically becomes equal to (\ref{suinfatorus}), which is 
homomorphic to $ADiff(T^{(2)})$. We will later explain what we mean by asymptotic equality.

It is clear that in (\ref{suinftorustriangint}) there is only $N-1$ independent structure 
coefficients, because they are invariant under shifting of indices:
\be
(m_1, m_2) \rightarrow (m_1 + lN, ~ m_2 + qN), \quad l, ~ q ~ \in \mathbb{Z} \label{shiftsymm}
\ee
This symmetry is in addition to modular transformation 
$\mathbf{m} \rightarrow \mathbf{m'} = U \mathbf{m}$, where $U$ is a $2 \times 2$ unimodular 
matrix with integer elements. Owing to these symmetries, the space of indices that characterize 
the algebra (\ref{suinftorustriangint}) can be considered as a 2D lattice of cell size $N$ in 
which boundaries are pairwise identified, that is $m_1 + lN \eqcirc m_1 + l'N$ and 
$m_2 + qN \eqcirc m_2 + q'N$. Here, the symbol $\eqcirc$ means periodic identification. 
This structure corresponds to flat space representation of high genus compact Riemann 2D 
surfaces. Therefore, when $N \rightarrow \infty$ the algebra (\ref{suinftorustriangint}), 
which owing to (\ref{kintop}) is associated to $\suinf$, is also homomorphic to $ADiff$ of 
compact orientable Riemann surfaces of any genus. Moreover, indices $(m_1 + lN, m_2 + qN)$ 
enumerate independent cocycles of these surfaces. A detailed proof of the homomorphism of 
$ADiff$ of any compact orientable 2D Riemann surface with $SU(N \rightarrow \infty)$ for $N$ 
countable is given in~\cite{suninfrep0}. We call {\it diffeo-surface} those surfaces which their 
$ADiff$ is at least asymptotically homomorphic to $\suinf$.
 
The algebra (\ref{suinftorustriangint}) does not include any information about diffeo-surfaces 
to which the corresponding $SU(N \rightarrow \infty)$ is homomorphic for a given $N$, because 
the structure coefficients of this algebra are independent of cocycle indices $(l,q)$ in 
(\ref{shiftsymm}). On the other hand, diffeomorphism cannot transfer Riemann surfaces with 
different genera to each others. Therefore, diffeo-surfaces with different genus should be 
considered as different representations of $\suinf$ symmetry.

\section{No exact description of sine algebra on a 2D surface}  \label{app:sinalgebra}
In~\cite{suinftriang0} it is claimed that the following operators generate the algebra 
(\ref{suinftorustriang}):
\be
\hK_{\mathbf{m}} \equiv (\frac{ir}{2}) \exp (i \mathbf{m}.\mathbf{x}) ~ \hG_{\mathbf{m}} (k) 
\quad , \quad \hG_{\mathbf{m}} (k) \equiv \exp (k ~ \mathbf{m} \times \partial_{\mathbf{x}})  
\label {kgen}
\ee

It is straightforward to show that:
\bea
&& [\exp (i \mathbf{m}.\mathbf{x}), \hG_{\mathbf{n}} (k)] = -\exp (i \mathbf{m}.\mathbf{x}) 
\exp (ik ~ \mathbf{m} \times \mathbf{n}) \label{gexpcommut} \\
&& [\hK_{\mathbf{m}},\hK_{\mathbf{n}}] = (\frac{ir}{2})^2 \exp (i (\mathbf{m} + \mathbf{n})) 
\biggl (\exp (ik ~ \mathbf{m} \times \mathbf{n}) \hG_{\mathbf{n}} (k) - \exp (-ik ~ 
\mathbf{m} \times \mathbf{n}) \hG_{\mathbf{m}} (k) \biggr )  \label{kkcommutx}
\eea
Only when $k \rightarrow 0$ and $|\mathbf{m}|, ~ |\mathbf{n}| \gg 0$, the r.h.s. of 
(\ref{kkcommutx}) approximately corresponds to the r.h.s. of (\ref{suinftorustriang}), and for 
all other values of $k$, the r.h.s. of (\ref{kkcommutx}) and (\ref{suinftorustriang}) are 
different. Thus, question arises whether another expression for $\hK_{\mathbf{m}}$ as a 
functional of $\mathbf{x}$ and $\partial_{\mathbf{x}}$ can be found such that their commutations 
exactly corresponds to (\ref{suinftorustriang}). Here we prove that there is no such 
expression.

Consider $\hK_{\mathbf{m}} = f (\mathbf{x}, \mathbf{m} \times \partial_{\mathbf{x}})$. As 
$\mathbf{x}$ and $\partial_{\mathbf{x}}$ do not commute, an analytical expression for $f$ is 
ambiguous and needs an ordering. Giving the fact that at this stage there is no constraint on 
$f$, without loss of generality we can assume 
$\hK_{\mathbf{m}} = f_1 (\mathbf{x}) f_2(\mathbf{m} \times \partial_{\mathbf{x}}))$. The 
functions $f_1$ and $f_2$ can be separately Fourier expanded:
\bea
&& f_1 (\mathbf{x}) = \sum_{\mathbf{s}} F_1 (\mathbf{s}) \exp (i \mathbf{s}.\mathbf{x}), 
\quad \quad  \sum_q F_2 (q) \exp (i q ~ \mathbf{m} \times \partial_{\mathbf{x}})  
\label{f1f2fourier} \\
&& \hK_{\mathbf{m}} = \sum_{\mathbf{s}, q} F_1 (\mathbf{s}) F_2 (q) \exp (i \mathbf{s}.\mathbf{x}) 
\exp (i q ~ \mathbf{m} \times \partial_{\mathbf{x}})  \label{kmfourier}
\eea
Notice that $\mathbf{s}$ and $q$ are in general continuous variables and can be also complex. 
Thus, sums in (\ref{f1f2fourier}-\ref{kmfourier}) can be rather integrals. Using these 
expansions, we find:
\bea
[\hK_{\mathbf{m}}, \hK_{\mathbf{n}}] & = & \sum_{\mathbf{s}, q, \mathbf{s'}, q'} F_1 (\mathbf{s}) F_2 (q) 
F_1 (\mathbf{s'}) F_2 (q') \exp(i (\mathbf{s} + \mathbf{s'}))   \nonumber \\ 
&& \biggl (\exp (-q ~ \mathbf{m} \times \mathbf{s'}) \exp (iq' ~ \mathbf{n} \times 
\partial_{\mathbf{x}}) - \exp (-q' ~ \mathbf{n} \times \mathbf{s}) \exp (iq ~ \mathbf{m} \times 
\partial_{\mathbf{x}}) \biggr )  \nonumber \\ 
& = & \sum_{\mathbf{s}, q, \mathbf{s'}, q'} F_1 (\mathbf{s}) F_2 (q) 
F_1 (\mathbf{s'}) F_2 (q') \exp(i (\mathbf{s} + \mathbf{s'}))   \nonumber \\ 
&& \biggl (\exp (-q ~ \mathbf{m} \times \mathbf{s'}) \hG_{\mathbf{n}} (iq') - 
\exp (-q' ~ \mathbf{n} \times \mathbf{s}) \hG_{\mathbf{m}} (iq) \biggr )  \label{kkcommutgen}
\eea
Comparison of (\ref{kkcommutgen}) with (\ref{kkcommutx}) shows that the latter corresponds to 
a special case where $F_2 \neq 0$ only for one mode $q = -ik$ and $\mathbf{s} = \mathbf{m}$. 
Therefore, giving the independence of Fourier modes, the more general case of the algebra 
presented in (\ref{kkcommutgen}) cannot become homomorphic to (\ref{suinftorustriang}) for any 
choice of the functions $f_1$ and $f_2$.

\section{Equivalence of amplitude factorization and algebraic criteria for compositeness} \label{app:ampfactor}
Product states satisfy the three criteria for division of a quantum system to multiple subsystems 
according to~\cite{sysdiv}: 
\begin{description}
\item {\bf Commuting operator blocks: } We assume that $A_{k_1 k_2 \cdots k_n}$ in 
(\ref{nbodystate}) is factorized to $m$ blocks of dimension $r_j, j=1, \cdots, m$ and:
\be
\sum_{j=1, \cdots, m} r_j = n   \label{ricond}
\ee
Each factor is associated to a subspace $\hm_j \subset \hm$ of the many-body system Hilbert 
space $\hm$. The set of parameters $\{k_{j,i}, i = 1, \cdots, r_j\}$ characterize vectors of 
$\hm_j$. Using the expansion (\ref{nbodystate}), a factorized state $|\Psi_f \rangle$ can be 
written as:
\bea
|\Psi_f \rangle &=& \sum_{\substack{j=1, \cdots, m; \\ \{k_{j,i}, ~ i= 1, \cdots, r_j\}}} 
A^{(1)}_{k_{1,1} k_{1,2} \cdots k_{1,r_1}} \cdots A^{(m)}_{k_{m,1} k_{m,2} \cdots k_{m,r_m}} \nonumber \\
&& |\Psi_1 (k_{1,1}, \cdots, k_{1,r_1}) \rangle \times \cdots \times |\Psi_m (k_{m,1}, \cdots, 
k_{m,r_m}) \rangle \nonumber \\
&=& \bigotimes_{j=1, \cdots, m} \biggl (\sum_{\{k_{j,i}, ~ i= 1, \cdots, r_j\}} 
A^{(j)}_{k_{j,1} k_{j,2}\cdots k_{j,r_j}} |\Psi_j (k_{j,1}, \cdots, k_{j,r_j}) \rangle \biggr ) 
\label{ampfactorstate}
\eea
Vectors $|\Psi_j \rangle$ may be in turn expandable with respect to a tensor basis depending 
on $k_{j,i}$:
\be
|\Psi_j (k_{j,1}, \cdots, k_{j,r_j}) \rangle \equiv |\psi_{k_{j,1}}\rangle \times 
|\psi_{k_{j,2}}\rangle \times \cdots  \label{psii}
\ee
and the set of $\{|\psi_{k_{j,i}}\rangle\}$ constitute a tensor product basis for $\hm$. But, 
by assumption amplitudes $A^{(j)}_{k_{j,1} k_{j,2}\cdots k_{j,r_j}}$ would not be further factorizable.
Moreover, (\ref{psii}) shows that for different $j$'s states $|\Psi_j \rangle$ do not share 
characteristic parameters $k_{j,i}$. Thus, subspaces $\hm_j$ are disjoint. This is the reason 
for the last equality in (\ref{ampfactorstate}) and shows that the factorization of 
amplitudes of a state with a tensor product basis leads to symmetry breaking and 
factorization of the Hilbert space: 
\be
\hm \supseteq \hm_1 \times \ldots \hm_m  \label{hilbertdecomp}
\ee
Therefore, $\forall \hA \in \bm[\hm_i]$ and $\hB \in \bm[\hm_j],~ i \neq j, [\hA,\hB] = 0$.
\item {\bf Locality: } When the classical spacetime coordinates $x$ are included in the set of 
parameters characterizing state of a quantum system, locality means that components of the 
density matrix become negligibly small for 
$|x - y| \rightarrow \infty, x,y \in \mathbb{R}^{(3+1)}$ for \footnote{This is a 
naive definition of {\it locality} for a quantum system, but sufficient for our purpose here. 
A more rigorous definition should consider von Neumann classification of the system, see 
e.g.~\cite{vonneumanalgebrarev} for a review.}. More generally, $x$ can be any characterizing 
parameter of the quantum state. Lack of locality means impossibility of tensor decomposition, 
even as an approximation. Locality condition is fulfilled if the factorized amplitudes are 
local. Endomorphism between characterizing parameters of factorized and unfactorized amplitudes 
ensures that if it is satisfied by the full basis, it is also satisfied by the factorized one.
\item {\bf Complementarity: } A map $\varphi: \hm \rightarrow \hm$ can be defined such that each 
factorized state $|\Psi_f \rangle$ is projected to an unfactorized state. The endomorphism condition 
is satisfied if:
\be 
\varphi (|\Psi_f \rangle) = |\Psi \rangle, \quad \quad A^{(1)}_{k_{1,1} k_{1,2} \cdots k_{1,r_1}} 
\cdots A^{(m)}_{k_{m,1} k_{m,2} \cdots k_{m,r_m}} = A_{k_1 k_2 \ldots k_n} \label{endo}
\ee
where $\{k_l\}$ characterizes a basis for the full Hilbert space $\hm$. Considering 
(\ref{ricond}), there is an endomorphism between parameter sets 
$\{k_{i,j}, ~ i=1, \cdots, m; ~ j=1, \cdots, r_i\}$ and $k_l, ~ l=1, \cdots; n$. Therefore, 
complementary condition is fulfilled by states with factorized amplitudes.
\end{description}

In \sqgr~the state of the Universe is not truly factorizable, because the components in a 
tensor product basis similar to (\ref{nbodybasis}) constitute the assumed global $\suinf$ 
symmetry of the Hilbert space $\hm_U$. Hence, any factorization of amplitudes can be reverted 
by a $\suinf$ transformation under which observables remain unchanged. Nonetheless, 
approximate factorization, symmetry breaking, and division to subsystem arise due to 
quantum fluctuations and locality of dynamics. 

\section{Homomorphism of $\suinf$ and $G \times \suinf$}  \label{app:gsuingcongex}
An explicit homomorphism between generators of $\suinf$ and $G \times \suinf$ 
can be established as the following. First consider the simplest case of $G = SU(2)$. Using 
torus basis (\ref{hlmdef}) for the definition of $\suinf$ generators $\hL_{\mathbf{m}}$, the 
following bijection is an example of possible maps between 
$\hat{\sigma}_i \times \hL_{\mathbf{m}}$, where $\hat{\sigma}_i, ~ i = 1, 2, 3$ are Pauli 
matrices, and $\hL_{\mathbf{m}}$:
\be
\Phi_{tor}: \{\hat{\sigma}_i \times \hL_{\mathbf{m}}\} \rightarrow \{\hL_{\mathbf{m}}\} ~ 
\biggl | ~ \Phi_{tor} (\hat{\sigma}_i \times \hL_{\mathbf{m}}) = \hL_{\mathbf{3m+i}}, \quad 
\mathbf{i} \equiv (i-1, i-1)  \label{su2maptorus}
\ee
In the sphere basis~\cite{suninfhoppthesis} (reviewed in Appendix D of~\cite{houriqmsymmgr}) 
we can use formal description of $\suinf$ generators $\hL(\theta, \phi)$ with respect to angular 
coordinates on sphere to define a bijection:
\bea
\Phi_{sph} &:& \{\hat{\sigma}_i \times \hL (\theta,\phi) \} \rightarrow 
\{\hL (\theta,\phi)\} ~ \biggl | ~ \Phi_{sph} \biggl (\hat{\sigma}_i \times 
\hL (\theta,\phi) \biggr) = \hL (\theta, \frac {\phi + 2\pi (i-1)}{3}), \nonumber \\
&& \theta = [0, \pi], ~ \phi = [0, 2\pi]  
\label{su2mapsphere}
\eea
This example can be easily extended to other finite rank $G$. Specifically, (\ref{su2maptorus}) 
can be generalized to:
\be
\Phi_{tor}: \{\hat{\lambda}_i \times \hL_{\mathbf{m}}\} \rightarrow \{\hL_{\mathbf{m}}\} ~ 
\biggl | ~ \Phi_{tor} (\hat{\lambda}_i \times \hL_{\mathbf{m}}) = \hL_{\mathbf{d_G m+i}}, \quad 
\mathbf{i} \equiv (i-1, i-1), \quad i = 1, \cdots, n_G  \label{gmaptorus}
\ee
where $n_G$ is the dimension of $G$, that is the number of generators $\hat{\lambda}_i$.

\section {Entanglement measures}  \label{app:entangle}
Entanglement of quantum systems can be quantified and several entanglement measures are 
proposed in the literature~\cite{qmaventropy,prodstate,qmnegativity,qmnegativity0}. Here we 
calculate entanglement entropy and negativity for unspecified subsystems of the 
Universe. These estimations would be useful for future application of \sqgr~to physical 
systems and processes, such as black holes and their information puzzle, inflation, 
particle production during reheating of the Universe, and decoherence of quantum systems 
by gravitational interaction.

\subsection{Entanglement entropy}  \label{app:ententropy}
The amount of entanglement is reflected in mutual information 
$I(G:\suinf) = S(\hvarrho_G) + S(\hvarrho_\infty) - S(\hvarrho_U)$, where 
$S (\hvarrho) \equiv -\tr (\hvarrho \ln \hvarrho)$ is the von Neumann entropy. As 
$\hvarrho_U$ is pure, $S(\hvarrho_U) = 0$. In this case $S(\hvarrho_G) = S(\hvarrho_\infty)$, 
see e.g.~\cite{qmaventropy}. This can be confirmed by explicit calculation using 
(\ref{gdensity}) and (\ref{infdensity}). We find:
\be
I(\hvarrho_G:\hvarrho_\infty) = 2 S(\hvarrho_G) = 2 S(\hvarrho_\infty) = 
-4 \sum_{\{k_G; (\eta, \zeta, \cdots)\}} ~ |A (k_G; \eta, \zeta, \cdots)|^2 
\ln |A (k_G; \eta, \zeta, \cdots)|  \label {rhoentropy}
\ee
This calculation also demonstrates that the reason for $S(\hvarrho_G) = S(\hvarrho_\infty)$ is 
the complementarity of $\hm_G$ and $\hm_\infty$ as subspaces of $\hm_U$. Using normalization 
and unitarity constraint (\ref {amptracecond}), we find that mutual information 
$I(\hvarrho_G:\hvarrho_\infty) \geqslant 0$, which demonstrates the entanglement of $\hvarrho_G$ 
and $\hvarrho_\infty$ states and corresponding subsystems. Moreover, as (\ref{rhoentropy}) 
shows, the entropy cannot be divided to {\it geometrical} - $\suinf$ related - and 
{\it internal}. Indeed, this is what the Einstein equation tells us. Gravitational effects 
represented by the curvature of spacetime is directly related to the state of matter - 
subsystem of the Universe - represented by their energy-momentum tensor, and vis-versa.

It is useful to compare the entanglement entropy (\ref{rhoentropy}) which depends on the 
degrees of freedom of both internal symmetry of the subsystem and its quantum gravity, with 
the entanglement entropy in QGR proposals based on the holography principle, such as 
AdS/CFT~\cite{adscft5,adscftrev,adscftentangle} and related models that interpret variation of 
entanglement as gravity~\cite{qgrentangle,qgrentangle0,qgrentangle1}. First of all, equality 
of the Bekenstein entropy in a bulk spacetime with entropy of a dual Conformal Field Theory 
(CFT) on a boundary is demonstrated only for (2+1)D bulk with Anti-de Sitter (AdS) geometry. 
Evidently, this setup does not correspond to the (3+1)D physical spacetime with a positive 
cosmological constant and FLRW geometry. Even accepting the conjecture of the existence of a 
dual CFT for any geometry, the spacetime must have at least a virtual boundary on which the 
dual CFT could be defined and used for calculation of the {\it geometrical} entropy. AdS 
includes such a boundary, but FLRW does not. Although a priori horizon can play the role of a 
null surface, as requested in covariant formulation of holography~\cite {stringvacuarev,greos}, 
it does not solve the issue of indivisibility of QFT's due to their type III von Neumann 
algebra, see e.g~\cite{vonneumanalgebrarev} for a review. None of such restrictions exists 
in \sqgr. 

\subsection{Entanglement negativity}  \label{app:qmnegativity}
Another entanglement measure frequently advocated in the literature is entanglement 
negativity or the related quantity called logarithmic 
negativity~\cite{qmnegativity,qmnegativity0}. The latter is a monotone but non-convex 
measure of entanglement~\cite{qmnegativity0}. For a density matrix $\hrho$ they are defined as:
\be
N(\hrho_\mg) \equiv \frac{|| \hrho_\mg^T ||_1 -1}{2}, \quad \quad LN(\hrho_g) \equiv 
\log || \hrho_\mg^T ||_1   \label{qmneg}
\ee
where for any operator $\hO$, $||\hO||_1 \equiv \tr |\hO| = \tr \sqrt{\hO^\dagger \hO}$, and 
$\hrho_\mg^T$ is the partial transpose of $\hrho$ over its subspace $\mg$. We remind that 
partial transpose preserves the trace. Therefore, $\hrho_\mg^T$ is also a density matrix. 
Entanglement negativity has the advantage of being purely quantum, in contrast to mutual 
information which is also defined for classical systems. In particular, it better reflects 
evolution of entanglement, see e.g.~\cite{qmentangnegex}.

The partial transpose of $\hvarrho_U$ in (\ref{gudensity}) with respect to $G$ subspace, that 
is the subspace generated by $\hrho_G$ is:
\be
\hvarrho_{U_G}^T =  \sum_{\substack {\{k_G, k'_G\} \\ (\eta, \zeta, \eta', \zeta', \cdots)\}}} 
A (k_G; \eta, \zeta, \cdots)  A^*(k'_G, \eta', \zeta', \cdots) ~ \hrho_G (k'_G , k_G) \times 
\hrho_\infty (\eta, \zeta, \eta', \zeta', \cdots) \label{gudensitytrans} 
\ee
It is straightforward to verify that $\hvarrho_{U_G}^{T\dagger} = \hvarrho_{U_G}^T$. Therefore, 
$\hvarrho_{U_G}^T$ is also a density operator and $\tr \hvarrho_{U_G}^T = 1$. Thus:
\be
||\hvarrho_{U_G}^T||_1 = \tr \sqrt{\hvarrho_{U_G}^{T\dagger} \hvarrho_{U_G}^T} = \tr |\hvarrho_{U_G}^T|
\ee
The necessary - but not sufficient - condition for separability of $\hvarrho_U$ is positivity 
of $\hvarrho_{U_G}^T$ eigen values~\cite{prodstate}. In this case 
$\tr |\hvarrho_{U_G}^T| = \tr \hvarrho_{U_G}^T = 1$ and negativity of $\hvarrho_U$ would be zero. 
However, explicit calculation shows that $(\hvarrho_{U_G}^T)^2 \neq \hvarrho_{U_G}^T$:
\bea
(\hvarrho_{U_G}^T)2 & = & \frac{1}{2} \biggl\{\mathlarger{\sum}_\Omega ~ 
\biggl (A (k_1; \eta_1, \zeta_1, \cdots) ~ A^*(k'_1; \eta'_1, \zeta'_1, \cdots) ~ 
A(k_2; \eta'_1, \zeta'_1, \cdots) ~ A^* (k_1; \eta'_2, \zeta'_2, \cdots) ~ \nonumber \\
&& \hrho_G (k'_1, k_2) \times \hrho_\infty (\eta'_1, \zeta'_1; \eta'_2, \zeta'_2, \cdots) \biggr ) + \nonumber \\
&& \mathlarger{\sum}_{\Omega'} ~ \biggl (A (k_2; \eta'_2, \zeta'_2, \cdots) ~ 
A^*(k'_2; \eta_1, \zeta_1, \cdots) ~ A(k_1; \eta_1, \zeta_1, \cdots) ~ 
A^* (k_2; \eta'_1, \zeta'_1, \cdots) ~ \nonumber \\
&& \hrho_G (k'_2, k_1) \times \hrho_\infty (\eta_2, \zeta_2, (\eta'_1, \zeta'_1, \cdots) \biggr ) 
\biggr\}  \label{gudensitytrans2} \\
\Omega & \equiv & \{k_1, k'_1, k_2\}; \{(\eta_1, \zeta_1, \cdots)\}; 
\{(\eta'_1, \zeta'_1, \cdots)\}; \{(\eta'_2, \zeta'_2, \cdots)\}; \label{indices} \\
\Omega' & \equiv & \{k_1, k_2, k'_2\}; \{(\eta_1, \zeta_1, \cdots)\}; 
\{(\eta'_1, \zeta'_1, \cdots)\}; \{(\eta_2, \zeta_2, \cdots)\}; \label{indicesp}
\eea
It is important to notice that tensor product basis $\hrho_G$ and $\hrho_\infty$ depend only 
on two sets of parameters $\{k\} and \{\eta, \zeta, \cdots\}$. There are, however, three sets of 
parameters in the summations in (\ref{gudensitytrans2}). They are indicated as $\Omega$ and 
$\Omega'$ in(\ref{indices}) and (\ref{indicesp}), respectively. For this reason, amplitudes 
in (\ref{gudensitytrans2}) cannot be factorized, unless $k$ and $(\eta, \zeta, \cdots)$ 
dependence of $A (k; \eta, \zeta, \cdots)$ can be factorized. However, the assumed $\suinf$ 
symmetry of the Universe mixes the two sectors and in general dependency on $k$ and 
$(\eta, \zeta, \cdots)$ of the amplitudes $A(k; \eta, \zeta, \cdots)$ are not separable. Thus, 
$(\hvarrho_{U_G}^T)^2 \neq \hvarrho_{U_G}$, the density matrix $\hvarrho_{U_G}$ is a mixed state, 
and is not separable. This means that  $\hvarrho_{U_G}$ may have negative eigen 
values~\cite{prodstate}, and $N(\hvarrho_U) > 0$, $LN(\hvarrho_U) > 0$. These results confirm 
the conclusion made from entanglement entropy and Proposition 1 in~\cite{hourisqgr} about 
entanglement of every subsystem to the rest of the Universe. 

\section{Conditions for perfect purification}  \label{app:purifcond}
Imposing perfect purification conditions (\ref{rhoginfconstraint}) does not guarantee that they 
would be satisfied in the framework of \sqgr~and construction of subsystems as explained in 
Sec. \ref{sec:gstate}. Necessary conditions for purification of a set of quantum states and 
faithfulness of such process are investigated in~\cite{qmpurifcond}. We apply their approach to 
mixed states such as (\ref{gdensity}) and (\ref{infdensity}) in \sqgr. 

The process of purifying a mixed state $\hrho \in \bm[\hm]$ can be considered as a map:
\be
\Uplambda: \bm[\hm] \rightarrow \bm[\hm \times \hm'] ~ \biggl | ~ \Uplambda (\hrho) = 
\hrho', \quad \hrho \in \bm[\hm], ~ \hrho' \in \bm[\hm \times \hm'], ~ \hrho'^2 = \hrho'. 
\label{purifmap}
\ee
In order to project a density matrix to a density matrix, the map $\Uplambda$ must be 
completely positive and trace preserving. Moreover, for a faithful purification $\Uplambda$ 
must be linear\footnote{Faithfulness means tracing over the auxiliary space used for the 
purification should recover the original density matrix. Thus, if $\hrho_i,~i=1,2$ can be 
faithfully purified by $\Uplambda$, so does $\hrho_1 + \hrho_2$. Moreover, following 
equalities are satisfied: $\tr_{aux} \Uplambda (\hrho_i) = \hrho_i$ and 
$\tr_{aux}\Uplambda(\hrho_1) + \tr_{aux}\Uplambda (\hrho_2) = \tr_{aux}\Uplambda (\hrho_1 + \hrho_2)$. 
The latter equality would be valid for any density matrix only if 
$\Uplambda (\hrho_1 + \hrho_2) = \Uplambda (\hrho_1) + \Uplambda (\hrho_2)$. Thus, 
$\Uplambda$ must be linear.}. Such maps can be obtained from appending a pure ancilla state 
$|\upchi \rangle \langle \upchi|$ to a mixed state $\hat{\upsigma}$ and performing a unitary 
rotation $U$ on the extended state. On the other hand, purifying process needs extending the 
state and its Hilbert space by an auxiliary state $\hat{\upomega}_{aux}$. A set of states are 
called {\it essentially pure} if the result of the above sets of operations are equal: 
\be
\hrho \times \upomega_{aux} = U (|\upchi \rangle \langle \upchi | \times \hat{\upsigma}) U^\dagger
\label{essenpurecond}
\ee
It is demonstrated~\cite{qmpurifcond} that an essentially pure set of states, and more generally 
orthogonal union of such sets have a perfect purifier. 

Here we show that the set of states $\hvarrho_G$ calculated in (\ref{gdensity}) are essentially 
pure and have a perfect purifier. Irrespective of the value of amplitudes $A_G$ in 
(\ref{gdensity}), the set of $\hvarrho_G$ states have the same general structure. Thus, it is 
enough to demonstrate essentially pureness for one of them. As we had traced out the $\suinf$ 
component of the tensor product basis $\hrho_G \times \hrho_\infty$ in (\ref{gdensity}) to obtain 
$\hvarrho_G$, the natural candidate for the auxiliary Hilbert space to append to $\hm_G$ is a 
space generated by the basis $\hrho_\infty (\eta, \zeta, \eta', \zeta', \cdots)$. We choose the 
pure state $|\upchi \rangle$ and the mixed state $\hat{\upsigma}$ belonging to a 
representation of $\suinf$ as the followings:
\bea
|\upchi \rangle & \equiv & \sum_{k_G} \chi (k_G) |k_G\rangle, \quad \quad \hat{\upchi}_G = 
|\upchi \rangle \langle \upchi |  \label{pureg}  \\
\hat{\upsigma} & = & \sum_{\{(\eta, \zeta, \cdots)\}, \{(\eta', \zeta', \cdots)\}} 
\upsigma (\eta, \zeta; \eta', \zeta'; \cdots) ~ \hrho_\infty (\eta, \zeta ; \eta', \zeta'; 
\cdots), \label{mixedinf} 
\eea
In Appendix \ref{app:gsuingcongex}~ we proved that $G \times \suinf \cong \suinf$. Thus,   
$\hat{\upchi}_G \times \hat{\upsigma} \in \bm[\hm_{G_\infty]} \cong \bm[\hm_U]$ and the unitary 
operator $U$, which must perform a global rotation is also in $\bm[\hm_{G_\infty]}$. The result 
of this rotation can be expanded as:
\be
U (\hat{\upchi}_G \times \hat{\upsigma}) U^\dagger = 
\sum_{\substack {\{k_G, k'_G \} \\ (\eta, \zeta, \eta', \zeta', \cdots)\}}} 
C (k_G, k'_G; \eta, \zeta; \eta', \zeta'; \cdots) ~ \hrho_G (k_G , k'_G) \times 
\hrho_\infty (\eta, \zeta, \eta', \zeta', \cdots),  \label{globalrotate} \\
\ee
Thus the r.h.s. of (\ref{essenpurecond}) for $\hvarrho_G$ is also in $\bm[\hm_{G_\infty]}$, and a 
minimal choice for $\hat{\upomega}_{aux}$ would be an operator in $\bm[\hm_\infty]$ expanded 
by the basis $\hrho_\infty$:
\be
\hat{\upomega}_{aux} = \sum_{\{(\eta, \zeta, \cdots)\}, \{(\eta', \zeta', \cdots)\}}  
\upomega (\eta, \zeta; \eta', \zeta'; \cdots) ~ \hrho_\infty (\eta, \zeta ; \eta', \zeta'; 
\cdots), \label{auxstate} 
\ee
The essentially pure condition (\ref{essenpurecond}) is satisfied for $\hvarrho_G$ and above 
choices of auxiliary Hilbert space and associated operators if:
\be
A_G (k_G; \eta, \zeta, \cdots) A^*_G (k'_G, \eta, \zeta, \cdots) 
\upomega (\eta, \zeta; \eta', \zeta'; \cdots) = C (k_G, k'_G; \eta, \zeta; \eta', \zeta'; \cdots), 
\quad \forall k_G, k'_G; \eta, \zeta; \eta', \zeta'; \cdots  \label{purifamp}
\ee
and the condition for perfect purification is satisfied if:
\be
\sum_{\{(\eta', \zeta', \cdots)\}} C (k_G, k'_G; \eta, \zeta; \eta', \zeta'; \cdots) = 
A_G (k_G; \eta, \zeta, \cdots) A^*_G (k'_G, \eta, \zeta, \cdots)  \label{opampunitary}
\ee
After summing (\ref{purifamp}) over $\{(\eta', \zeta', \cdots)\}$, we find the following 
constraint for $\upomega$:
\be
\sum_{\{(\eta', \zeta', \cdots)\}} \upomega (\eta, \zeta; \eta', \zeta'; \cdots) = 1  
\label{upomegacond}
\ee
This constraint on the auxiliary state $\upomega$ is independent of how $A_G$ amplitudes in 
$\hvarrho_G$ depend on the {\it external} parameters $(\eta, \zeta, \cdots)$. Therefore, this 
purification procedure is universal, and $\hat{\upomega}_{aux}$ can be chosen such that 
$\hvarrho_G \times \hat{\upomega}_{aux}$ be a faithful purification of $\hvarrho_G$. 

\section{$\mathbf{\suinf}$ Yang-Mills QFT on a background spacetime} \label{app:ymspacetime}
The $\suinf$-Yang-Mills theory on a background spacetime is first studied in~\cite{suninfym}. 
For a finite rank non-Abelian symmetry $G$ the gauge field is defined as: 
\be
A_\mu (x) \equiv \sum_a A_\mu^a T^a  \label{gaugedef}
\ee
where $x$ is a point in the background spacetime and $a$ is an index - usually an integer - 
enumerating symmetry generators $T^a$ of $G$ in a given representation. When the symmetry 
group is the infinite rank $\suinf$, the definition (\ref{gaugedef}) still applies with the 
difference that index $a$ consists of two independent continuous variables, because $\suinf 
\cong ADiff(D_2)$. For instance, in sphere basis $a \equiv (\theta, \phi) \in S^{(2)}$ are 
angular coordinates on a sphere. Generators of $\suinfa$ algebra can be formally described 
with respect to these parameters, see appendices of~\cite{houriqmsymmgr}. Alternatively, they 
can be decomposed to spherical harmonic functions $Y_{lm} (\theta,\phi)$, which satisfy Poisson 
bracket (algebra) in the l.h.s. of (\ref{suinfgendef}). In turn, Poisson algebra is homomorphic 
to $\suinfa$~\cite{suninfhoppthesis}. Therefore, 
$a \equiv \{(\theta,\phi); ~ (l \geqslant 0, |m| \leqslant l \}$. As we reviewed in Appendix 
\ref{app:suinfrev}, in torus basis generators are expressed as 2D harmonic oscillators 
depending on two continuous parameters $(x_1,x_2) \equiv \mathbf{x}$ and a 2D integer vector 
$(m_1,m_2) \equiv \mathbf{m}$. Thus, $a \equiv \{\mathbf{x}; \mathbf{m}\}$~\cite{suinftriang0}. 

Using sphere basis, which is used by~\cite{suninfym}, the $\suinf$ gauge field defined on a 
background spacetime is expressed as: 
\be
A_\mu(x; \theta, \phi) \equiv \sum_{lm} A_\mu^{lm}(x) \hL_{lm} (\theta, \phi) \label{suinfgauge}
\ee
The authors of~\cite{suninfym} call the 2D surface generated by $(\theta,\phi)$ 
{\it the internal space} to distinguish it from the background spacetime. It corresponds to 
what we have called {\it diffeo-surface} of $\suinf$ representation in the context of \sqgr.

\section{Quantum fields and density matrices} \label{app:fielddens}
In QFT's a field operator $\hphi$ acts on a multi-particle Fock space and depends on the 
parameters constituting a space $\Upxi$. The latter is usually - but not necessarily - the 
spacetime or its conjugate mode space. The field is often decomposed to creation and 
annihilation operators:
\bea
&& \hphi (y) \equiv \ha (y) + \ha^\dagger (y), \quad y \in \Upxi \label{fielddef}  \\ 
&& \ha^\dagger (y) |\emptyset\rangle = |y\rangle, \quad \langle \emptyset | \ha^\dagger = 0, \quad 
\ha (y) |\emptyset\rangle = 0, \quad \langle \emptyset | \ha = \langle y|, \quad 
[\ha (y), \ha^\dagger (y')] = 1.  \label{ceartannildef}
\eea
The third equality in (\ref{ceartannildef}) defines the vacuum state $|\emptyset\rangle$ 
and the first equality defines a single particle state at point $y$ of the parameter space. 
The commutation (or anti-commutation for fermions) is the quantization condition of the field. 
Notice that although creation and annihilation operators $\ha^\dagger$ and $\ha$, respectively, 
act on the multi-particle Fock space, their action is restricted to a single particle/subsystem. 
Therefore, $\ha (y), \ha^\dagger(y), \hphi (y) \in \bm[\hm]$, where $\hm$ is the single 
particle/subsystem Hilbert space parameterized by $\Upxi$. 

In a quantum mechanical view, the vacuum $|\emptyset\rangle$ can be identified with the null 
vector of the Hilbert space $\hm$, and $|y\rangle, ~ y\in \Xi$ with eigen vector of an 
operator, which with eigen value $y \in \Xi$. Therefore, a general 1-particle state 
$|\psi\rangle \in \hm$ and its density matrix $\hrho \in \bm [\hm]$ can be written as:
\bea
|\psi \rangle & = & \sum_y \psi (y) |y\rangle = \sum_y \psi (y) a^\dagger(y) |\emptyset\rangle, 
\quad \sum_y |\psi (y)|^2 = 1, \quad \langle y | y' \rangle = \delta (y-y')  \label{genstate} \\
\hrho & = & |\psi \rangle \langle \psi| = \frac{1}{2} \sum_{y,y'} [\psi (y) \psi^* (y') 
\ha^\dagger (y) |\emptyset\rangle \langle \emptyset | \ha(y') + C.C.]  \label{gendens}
\eea
The description of density matrix with respect to creation and annihilation operators 
demonstrates that fields and density matrices are different way of presenting state of a 
quantum particle/system, and one can be easily deduced from the other. The formulation of 
dynamics using a path integral defined on the configuration space with fields and their 
derivatives as generalized coordinates has its analogous with density matrices defined 
on the Hilbert space, see e.g.~\cite{qmopenbook} (Chapter 6) for details.

\end{document}